\documentclass[a4paper,fleqn,usenatbib,useAMS]{mnras}

\usepackage{graphicx}	
\usepackage{amsmath}	
\usepackage{amssymb}	
\usepackage{multicol}        
\usepackage{bm}		
\usepackage{pdflscape}	
\usepackage{amssymb}
\usepackage{tablefootnote}

\usepackage[T1]{fontenc}
\usepackage{ae,aecompl}

\usepackage{mathptmx}

\title[A search for TTV and orbital decay in WASP-46b]{A search for transit timing variations and orbital decay in WASP-46b\thanks{This work is partially based on observations obtained with the 1.54-m telescope at Estaci\'on Astrof\'isica de Bosque Alegre dependent on the National University of C\'ordoba, Argentina.}\thanks{Partially based on observations made with ESO Telescopes at the La Silla Paranal Observatory under programme ID 60.A-9022(A).}}

\author[R. Petrucci et al.]{R. Petrucci$^{1,3}$\thanks{Contact e-mail: \href{mailto:romina@oac.unc.edu.ar}{romina@oac.unc.edu.ar}}\thanks{Visiting Astronomer, Complejo Astron\'omico El Leoncito operated under agreement between the 
Consejo Nacional de Investigaciones Cient\'ificas y T\'ecnicas de la Rep\'ublica Argentina and
the National Universities of La Plata, C\'ordoba and San Juan.},
E. Jofr\'{e}$^{1,3}$, L. V. Ferrero$^{1,3}$, V. C\'uneo$^{1,3}$, L. Saker$^{1,3}$, F. Lovos$^{1,3}$, \newauthor{M. G\'{o}mez$^{1,3}$ and P. Mauas$^{2,3}$}\\
\\
$^{1}$Universidad Nacional de C\'ordoba, Observatorio Astron\'{o}mico, Laprida 854, X5000BGR, C\'{o}rdoba, Argentina \\
$^{2}$ Instituto de Astronom\'{i}a y F\'{i}sica del Espacio (IAFE, CONICET-UBA), Av. Inte. G\"{u}iraldes 2620, C1428ZAA, Buenos Aires, Argentina\\
$^{3}$Consejo Nacional de Investigaciones Cient\'{i}ficas y T\'{e}cnicas (CONICET), Argentina}

\date{Last updated 2015 May 22; in original form 2013 September 5}

\pubyear{2016}

\begin{document}
\label{firstpage}
\pagerange{\pageref{firstpage}--\pageref{lastpage}}
\maketitle

\begin{abstract}
We present 12 new transit observations of the exoplanet WASP-46b obtained with the 1.54-m telescope at Estaci\'on Astrof\'isica de Bosque Alegre (EABA, Argentina) and the 0.40-m Horacio Ghielmetti and 2.15-m Jorge Sahade telescopes at Complejo Astron\'omico El Leoncito (CASLEO, Argentina). 
We analyse them together with 37 light curves from the literature to re-determine the physical parameters and search for additional planets via transit timing variations (TTVs).  
We consider the 31 transits with uncertainties in their mid-transit times ($e_\mathrm{T_{0}}$) $<$ 1 minute, to perform the first homogeneous study of TTVs for the system, finding a dispersion of $\sigma= 1.66$ minutes over a 6 year baseline. Since no periodic variations are found, our interpretation for this relatively high value of $\sigma$ is that the stellar activity could be affecting the measured mid-transit times. This value of dispersion allows us to rule out the presence of additional bodies with masses larger than 2.3, 4.6, 7, and 9.3 $M_{\mathrm{\earth}}$ at the first-order mean-motion resonances 2:1, 3:2, 4:3, and 5:4 with the transiting planet, respectively.
Despite the 6 year baseline and a typical light curve precision of $2 \times 10^{-3}$, we find that we cannot significantly demonstrate a slow decrease of the orbital period of WASP-46b. We place a lower limit of Q$_{\star}$ > $7 \times 10^{3}$ on the tidal quality factor and determine that an additional 6 year baseline is required to rule out Q$_{\star}$ < $10^{5}$.

\end{abstract}






\section{Introduction}

WASP-46b is a Hot Jupiter-like planet orbiting a main-sequence G6 star (V$=12.9$, K$=11.4$), discovered by \citet{anderson} from data of the WASP photometric survey \citep{pollacco} taken during the years 2008 and 2009. \citet{anderson} estimated a planet mass of $M_\mathrm{P}=2.101 \pm 0.073$ $M_\mathrm{J}$ and a planetary radius of $R_\mathrm{P}=1.310 \pm 0.051$ $R_\mathrm{J}$ from two transits observed with the 1.2-m Euler and 0.6-m TRAPPIST telescopes and 16 radial velocity measurements taken with the CORALIE spectrograph. The spectroscopic observations confirm the period of 1.43 days found from the photometric data. The detection of weak emission in the Ca II H+K lines of the CORALIE spectra and a rotational modulation of 16 $\pm$ 1 days found in the WASP data confirm that WASP-46 is an active star. Combining this photometric information with the spectroscopically determined rotation velocity, \citet{anderson} inferred an inclination for the stellar spin axis of 41$^{\circ}$ with respect to the sky plane. They also found an inconsistency between the age of a few Gyr estimated from the lithium abundance and a gyrochronological age of 0.9-1.4 Gyr calculated from the stellar rotation period. 

Recently, \citet{maxted} used two improved methods to estimate the gyrochronological and isochronal ages of 28 transiting exoplanets. For about half the sample, including WASP-46b, they confirmed the discrepancy found by \citet{anderson} between the age determined by the stellar rotation period and the one obtained by the isochrone fitting. Although there is still no conclusive evidence, the authors suggest as a possible explanation that tidal interaction between the star and the planet has produced a transfer of angular momentum from the planetary orbit to the rotation of the star. This increase in the stellar spin makes the star appear younger than it really is, causing a smaller value for the gyrochronological age than the one measured by using isochrones. 

\citet{chen} observed one secondary eclipse in the $g'$, $r'$, $i'$, $z'$, J, H, and K bands simultaneously using the GROND instrument mounted on the MPG/ESO 2.2 m telescope at La Silla in Chile.  They detected thermal emission in the J, H, and K bands. The brightness temperatures resulting from these measurements are consistent with a very poor heat redistribution efficiency in the atmosphere of WASP-46b. Also, \citet{zhou} reported the detection of two full secondary eclipses of WASP-46b in the near IR band K$_{S}$ with the IRIS2 infrared camera on the 3.9-m Anglo-Australian Telescope. For both eclipses, they measured depth values consistent with the result previously obtained by \citet{chen}. 

\citet{kjurk} obtained one complete transit of WASP-46b with a 0.40-m telescope and determined photometric parameters in agreement with those measured by \citet{anderson}. Subsequently, \citet{ciceri} observed 10 primary transits of this exoplanet with telescopes of different sizes, ranging from 1.2- to 3.58-m. They determined a slightly lower and more precise value for the planetary radius ($R_\mathrm{P}=1.189 \pm 0.037$ $R_\mathrm{J}$) than the one reported by \citet{anderson}. This result implies that the planet's density is larger than initially thought. These authors also performed the first transit timing variation (TTV) study for this exoplanet. However, the data were not analysed homogeneously since they considered not only the measurements of their mid-transit times but also several values of $T_\mathrm{0}$ directly extracted from the literature. Although their results indicate that a linear ephemeris is not a good fit to the observations, \citet{ciceri} did not find any periodic variation and discarded the presence of a third body gravitationally bound to the system. However, they pointed out the need to acquire more precise mid-transit times and to perform a more homogeneous analysis of these data to firmly establish if there are TTVs or not. Finally, they reported a small difference in the planetary radius measured in the $i'$ and $z'$ bands, that could indicate the presence of water vapor at $\lambda \sim 920$ nm and the absence of potassium at $\lambda \sim 770$ nm. 

The work by  \citet{ciceri} is the only TTVs study of WASP-46b. However, their results are not conclusive, perhaps due to the fact that not all the mid-transit times used to carry out the analysis were obtained applying the same fitting procedure and error treatment, which means the study is not fully homogeneous.
Taking this into account, in this paper, we perform the first homogeneous TTVs study from the analysis of literature data and 12 new transit light curves of WASP-46b collected from three different telescopes. 
Furthermore, since a direct implication of the results obtained by \citet{maxted} for WASP-46 is that the planetary orbit would be shrinking, we also investigate the possibility of orbital decay in the system.

This paper is organised as follows: In Section 2 we present our observations and briefly describe the data reduction. In Section 3 we obtain the fundamental stellar parameters and chemical abundances, and we determine the photometric and physical parameters of the system. We also compare the computed values with the results obtained by other authors. In Section 4 we present our study of transit timing variations and the results of the search for orbital decay. The analysis of long-term variations on depth ($k$) and orbital inclination ($i$) is also described here. Finally, in Section 5, we present a summary and the conclusions.

\section{OBSERVATIONS AND DATA REDUCTION}

We observed 12 new transits of WASP-46b between July 2012 and July 2016. In Table 1 we present a log of our observations.
Five complete\footnote{We use the word ``complete'' to refer to ``full transit coverage''.} light curves were obtained with the Horacio Ghielmetti telescope (THG) located at Complejo Astron\'omico El Leoncito (CASLEO). The THG is an MEADE-RCX 400 telescope with a 0.40-m primary mirror, currently equipped with an  
Apogee Alta U16M camera and Johnson UBVRI filters. Due to a serious electric damage, the U16M was not available for the transit observed during the night of 2012, July 22. Therefore, only in this case, we used a different camera, an Apogee Alta U8300 with 3326$\times$2504 5 $\mu$m-size pixels, a scale of 0.32 arcsec per pixel, and a FoV=19'$\times$14'. The other 4 light curves were obtained with the Apogee Alta U16M camera with 4096$\times$4096 9 $\mu$m-size pixels, FoV=49'$\times$49', and a scale of 0.57 arcsec per pixel. 
For each night we also took 10 bias and 10 dark frames. Sky flat-field images were not taken since we previously found that flat-fielding correction causes unwanted errors in the photometric data \citep{petrucci13}. Averaged bias and median-combined bias-corrected dark frames were subtracted from science images using standard IRAF\footnote{IRAF is distributed by the National Optical Astronomy Observatories, which are operated by the Association of Universities for Research in Astronomy, Inc., under cooperative agreement with the National Science Foundation.} routines.

Other 6 light curves were obtained with the 1.54-m telescope located at Estaci\'on Astrof\'isica de Bosque Alegre (EABA). This telescope, operated in Newtonian focus, is currently equipped with Johnson UBVRI filters and a 3070$\times$2048 9 $\mu$m-size pixels Apogee Alta U9 camera, which provides a scale of 0.25 arcsec per pixel and a FoV=8'$\times$12'. We used this configuration to obtain 5 light curves. However, as a consequence of an electric flaw, the U9 CCD had to be replaced, and for the transit observed during the night of 2016, June 10 we employed a different camera, an Apogee Alta F16M with 4096$\times$4096 9 $\mu$m-size pixels, a scale of 0.25 arcsec per pixel, and a FoV=16.8'$\times$16.8'. Except for one transit, all the light curves were observed complete with their four points of contact visible. The partial transit was acquired during the night of 2014, August 22. In this case, the presence of clouds prevented us from obtaining data before the ingress and between the first and second contact points. A total of 10 bias, 8 dark, and 15 dome flat-field frames were taken for each observation. We corrected the EABA images for bias and dark applying the same procedure adopted for the THG ones. These CCD images were divided by the master flat generated as the median combined bias- and dark-corrected flats in the corresponding band. 

The remaining transit was obtained during the night of 2016, July 30 with the 2.15-m Jorge Sahade telescope at CASLEO. In this case, we used the Roper Scientific camera with 2048$\times$2048 13.5 $\mu$m-size pixels, a Johnson R filter, and a focal reducer which provided a circular FoV of 9' radius at a plate scale of 0.45 arcsec per pixel. We took 10 bias and 10 dome flat-field frames. Since the dark current level is quite low ($<$ 1 e$^{-}$/hr/px) dark frames were not taken. We corrected the CASLEO images for bias and then divided by the master flat generated as the median combined bias-corrected flats applying the procedure previously explained.

Contrary to the results obtained for the THG images, in both, CASLEO and EABA images, the flat-fielding correction produces light curves with dispersion values equal or, most of the time, smaller than those from light curves achieved without applying this correction.

Integration times ranging from 10 to 60 seconds and different CCD bin sizes for the science images were chosen depending on seeing, airmass, and atmospheric conditions during the night. The transits' observations were mainly carried out in the $R$ filter to decrease the effects of limb-darkening \citep{mallen}. However, one transit was observed with no filter (clear), to increase the signal-to-noise ratio without losing temporal resolution.
For all the observations we specially checked the focus of the telescope, to avoid any contamination from the faint star located near WASP-46 at a separation of 17.4" \citep{anderson}. Central times of the images were recorded in Heliocentric
Julian Date based on Coordinated Universal Time ($HJD_\mathrm{UTC}$).

Instrumental magnitudes (m$_\mathrm{ins}$) were measured through aperture photometry by using the FOTOMCAp code \citep{petrucci16}. This is a new quasi-automatic
program, written in IRAF, developed to determine precise instrumental magnitudes by applying the method of aperture correction \citep{howell, stetson}. FOTOMCAp has been demonstrated to significantly improve the results obtained with our previous code FOTOMCC \citep{petrucci13, petrucci15}, allowing not only a decrease in the standard deviation of the light curves but also an increase in the number of useful images to perform the aperture photometry. The main difference between both codes is that FOTOMCC uses the growth curves technique\footnote{It consists in the determination of the stellar flux variation as a function of aperture radius, in which the adopted magnitude will be given by the aperture at which the total flux of the source stops increasing.} to determine instrumental magnitudes, that sometimes can lead to an incorrect determination of m$_\mathrm{ins}$ as a consequence of variable background behind the source, incorrect sky background subtraction, or some other errors that affect fainter objects more than brighter ones. FOTOMCAp overcomes these issues by using the method of aperture correction in which a constant is calculated for each image as the median value of the magnitudes obtained from the aperture that allows the highest signal-to-noise ratio (m$_\mathrm{SN}$) minus those from the growth curves technique (m$_\mathrm{CC}$) for the brightest stars (the range of tested aperture radii is specified in Table 1). Then, the instrumental magnitudes of all the stars are computed by adding this constant or aperture correction to the m$_\mathrm{SN}$ determined for each star. Differential magnitudes were obtained through the procedure explained in \citet{petrucci13}, considering as comparison one or several stars of the same field with no indication of variability. It was not possible to use the same reference stars for all the transits because, as mentioned before, observations were carried out with several CCD cameras providing different fields of view. This fact prevented us from taking images of the exactly same stellar field and hence the same comparison stars for all the transits. However, in order to achieve the best light curve for each night, we selected as reference stars those which minimized the scatter in the resulting transit. We present our 12 transits with their best-fittings in Fig. 1 and the photometric data in Table 2.

\begin{table*}
\caption{Log of our observations}             
\label{table:1}      
\centering          
\begin{tabular}{lccccccccc}     
\hline\hline       
Date & Telescope & Camera & Filter & Bin-size & X & Exposure-Time (s) & N$_{\mathrm{obs}}$ & $\sigma$(mag) & Aperture radii (px)\\
\hline                    
2012 Jul 22  & 0.40-m THG & U8300 & clear & 4x4  & 1.10 $\rightarrow$ 1.51 & 20  & 707  &  0.0095 & 1-25 \\
2013 Aug 11  & 0.40-m THG & U16M & $R$ & 2x2  & 1.53 $\rightarrow$ 1.09 & 60 & 215 & 0.0085 & 1-25\\
2013 Aug 29  & 1.54-m EABA & U9 & $R$ & 2x2  & 1.11 $\rightarrow$ 1.53 & 50 & 204 & 0.0025  & 1-20\\
2013 Oct 10  & 0.40-m THG & U16M & $V$ & 2x2  & 1.14 $\rightarrow$ 1.79 & 50 & 209 & 0.0074 & 1-25\\
2014 Jun 30  & 0.40-m THG & U16M & $R$ & 2x2  & 1.25 $\rightarrow$ 1.09 $\rightarrow$ 1.26 & 50 & 314  & 0.0103 & 1-5\\
2014 Jul 23  & 1.54-m EABA & U9 & $R$ & 2x2  & 1.11 $\rightarrow$ 1.09 $\rightarrow$ 1.26 & 10 & 809 & 0.0034 & 1-20\\
2014 Aug 22  & 1.54-m EABA & U9 & $R$ & 2x2  &  1.09 $\rightarrow$ 2.06  & 55 & 225 &  0.0034 & 1-15 \\
2014 Sep 14  & 0.40-m THG & U16M & $R$ & 2x2  & 1.09 $\rightarrow$ 1.40  & 50 & 184  &  0.0046  & 1-25\\
2014 Oct 17  & 1.54-m EABA & U9 & $R$ & 2x2  & 1.09 $\rightarrow$ 1.54 & 20 & 601 & 0.0050 & 1-15\\
2015 Sep 08  & 1.54-m EABA & U9 & $R$ & 2x2  & 1.13 $\rightarrow$ 1.09 $\rightarrow$ 2.00 & 40, 50 & 480 & 0.0060 & 1-20\\
2016 Jun 10  & 1.54-m EABA & F16M & $R$ & 2x2  & 1.86 $\rightarrow$ 1.09  & 50 & 298  & 0.0028 & 1-25 \\
2016 Jul 30  & 2.15-m CASLEO & Roper & $R$ & 2x2  & 1.10 $\rightarrow$ 1.48  & 40  & 207  & 0.0027 & 1-12\\
\hline                  
\end{tabular}

Note: Date is given for the beginning of the transit, X is the
airmass change during the observation, N$_{\mathrm{obs}}$ is the number of useful exposures\tablefootnote{Some images had to be considered non useful because they were very shifted as a consequence of telescope movements to correct the position of the field, or because the stars showed a very low number of ADUs due to the presence of passing clouds.}, $\sigma$ is the standard deviation of the out-of-transit data-points, and Aperture radii is the range of aperture sizes in pixels tested by FOTOMCAp.
\end{table*}

\begin{figure*}
  \centering
 \includegraphics[width=.8\textwidth]{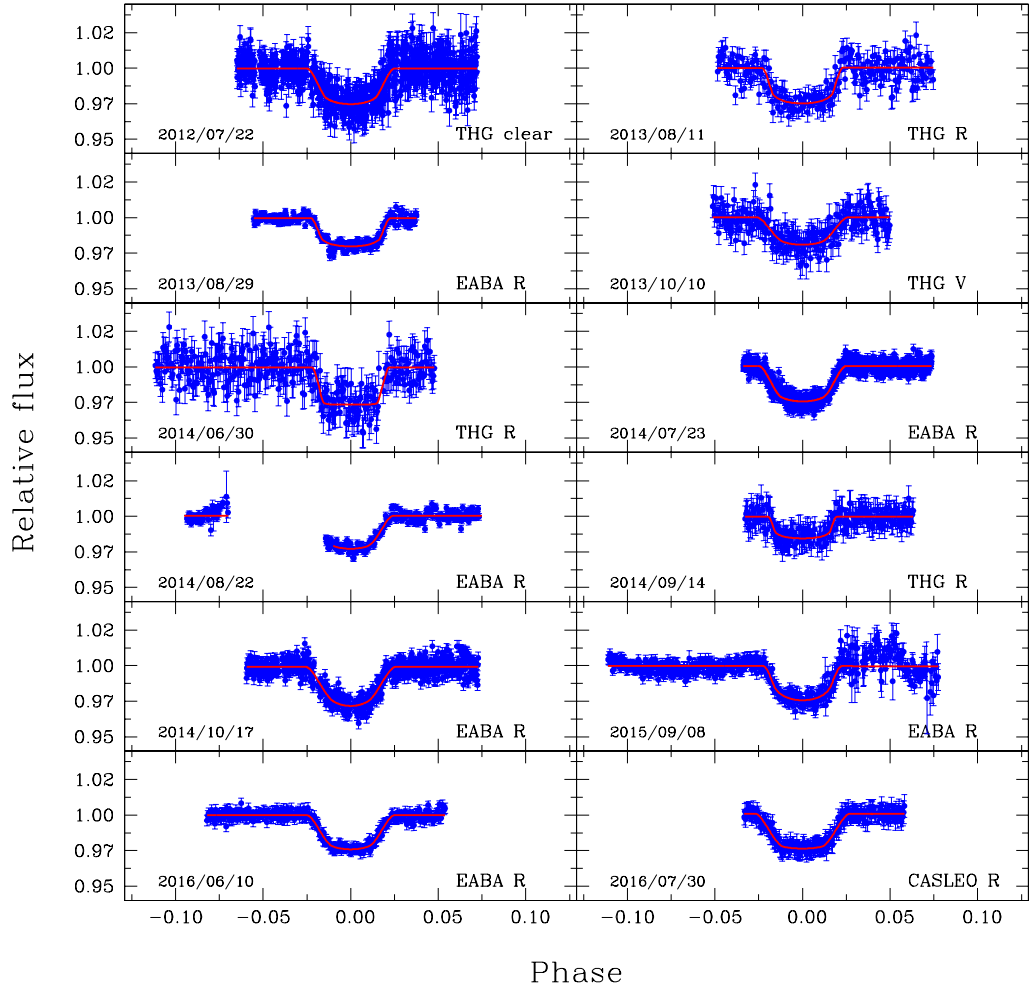}
   \caption{New transit observations presented in this work. Photometric data and their error bars (where errors are the measured ones) are indicated in blue, while the best-fittings are marked in solid lines. For each transit the observation date, observatory, and filter are also pointed out.}
     \label{FigVibStab}
\end{figure*}

\begin{table}
\caption{Photometry of WASP-46 obtained in this work. This table is available in its entirety in the online journal. A portion is shown here for guidance}             
\label{table:2}      
\centering          
\begin{tabular}{lccc}     
\hline\hline       
Telescope	& \textit{BJD}$_\mathrm{TDB}$ & Relative Flux & $\sigma_\mathrm{flux}$ \\
\hline      
0.40-m THG & 2456131.721802 & 1.005 & 0.008\\
0.40-m THG & 2456131.722057 & 0.994 & 0.008\\
0.40-m THG & 2456131.722312 & 1.004 & 0.008\\
0.40-m THG & 2456131.722566 & 0.999 & 0.008\\
0.40-m THG & 2456131.722821 & 1.001 & 0.008\\
0.40-m THG & 2456131.723076 & 1.006 & 0.008\\
0.40-m THG & 2456131.723330 & 0.999 & 0.008\\
0.40-m THG & 2456131.723585 & 0.995 & 0.008\\ 
             ...	&	...	& ... & ....\\
\hline                  
\end{tabular} 
\end{table}

\subsection{Literature and public data}
We supplemented our 12 transits of WASP-46b with 37 light curves available in the literature and public databases. We used 2 transits from \citet{anderson}: one observed in an $I+z'$ filter with the 0.60-m TRAPPIST telescope, and the other one acquired with the 1.2-m Euler-Swiss telescope in a Gunn $r'$ filter. Another 20 light curves from \citet{ciceri} were kindly provided to us by the authors. Among them, three were observed with the 1.54-m Danish Telescope using a Bessell $R$ filter, one with the 3.58-m New Technology Telescope (NTT) in a Gunn $g'$ filter, two with the 1.2-m Euler-Swiss telescope employing a Gunn $r'$ filter, three which were simultaneously observed in the four optical bands $g'r'i'z'$ with the GROND instrument on the 2.2-m MPG telescope, and finally one transit also obtained with GROND but only in the $g'z'$ filters. The remaining light curves were extracted from the Exoplanet Transit Database\footnote{The Exoplanet Transit Database (ETD) can be found at:
http://var2.astro.cz/ETD/credit.php; see also TRESCA at http://var2.astro.cz/EN/tresca/index.php} \citep{poddany}. We only included 15 complete and clearly visible transits of which 14 were observed without filter and the other one in the $R$ filter. Some of these light curves do not include the photometric errors. In those cases, we adopted as error the standard deviation of the out-of-transit data points. These observations were obtained with telescopes whose primary mirrors range from 0.25- to 1.54-m. 

\section{DETERMINATION OF THE SYSTEM'S PARAMETERS}

\subsection{Fundamental parameters and chemical abundances}

In order to derive the spectroscopic fundamental parameters ($T_\mathrm{eff}$, $\log g$, v$_\mathrm{turb}$, [Fe/H]) of WASP-46, we obtained 5 UVES high-resolution spectra from the ESO archive\footnote{http://archive.eso.org/cms.html} to produce a single spectrum (Fig. 2) with an average signal-to-noise ratio of  S/N $\sim$ 110 (around 6000 $\mathring{A}$). We employed the classical procedure, as previously described in \citet{jofre15a, jofre15b, petrucci13}. Briefly, fundamental parameters are computed from the equivalent widths (EWs) of iron lines (\ion{Fe}{I} and \ion{Fe}{II}) by imposing excitation and ionization equilibrium and the independence between abundances and EWs, using the FUNDPAR program \citep{saffe}. FUNDPAR employs the MOOG code \citep{sneden} and ATLAS9 1D local thermodynamic equilibrium (LTE) model atmospheres \citep{kurucz}. The resulting parameters, along with their statistical uncertainties, are listed in Table 3. Intrinsic uncertainties are based on the scatter of the individual iron abundances from each individual line and the standard deviations in the slopes of the least-squares fits of iron abundances with reduced EWs, excitation, and ionization potential \citep{gonzalez}. Overall, our parameters are consistent with those reported in the discovery paper \citep{anderson}, however, our $T_\mathrm{eff}$ value is $\sim$160 K warmer. This discrepancy, within its quoted error, might be related not only to the different technique performed by Anderson et al. to obtain $T_\mathrm{eff}$ (spectral synthesis of the H$_{\alpha}$ line) but also due to the higher signal-to-noise of our UVES final spectrum compared with their CORALIE spectra of S/N $\sim$ 50. Finally, we also computed the chemical abundances of 14 elements (Na, Mg, Al, Si, Ca, Sc, Ti, V, Cr, Mn, Co, Ni, Y, Ba) from the EWs of several unblended lines using the MOOG program (abfind driver) as in \citet{jofre15a}. The computed abundances, relative to the solar values of \citet{anders}, along with their dispersions around the mean are also included in Table 3. Our [X/H] values are consistent, within the errors, with those reported by Anderson et al. However, our abundances for all elements (except Mg and Sc) are systematically larger than those of Anderson et al. by $\sim$0.06 dex on average. The discrepancies here could be caused by the use of different line list and/or higher errors in the determination of the EWs due to the differences in the quality of the used spectra.  

\begin{figure*}
   \centering
   \includegraphics[width=.6\textwidth]{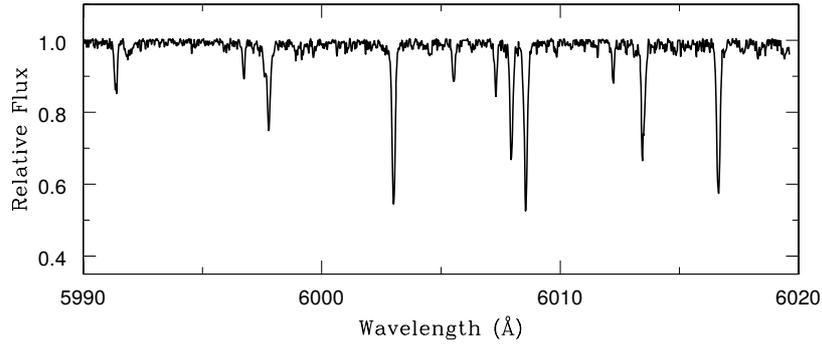}
      \caption{Observed UVES spectrum of WASP-46 in a narrow range around 6000 $\mathring{A}$ showing several metal lines. This spectrum was used to derive the fundamental stellar parameters and chemical abundances listed in Table 3.}
         \label{FigVibStab}
   \end{figure*}

\begin{table}
\caption{Fundamental parameters and chemical abundances of WASP-46 derived in this work from UVES spectra}             
\label{table:3}      
\centering          
\begin{tabular}{lc}     
\hline\hline       
Parameter (unit)	&	Value $\pm$ $\sigma$*\\
\hline                    
$T_\mathrm{eff}$ (K)	&	5761 	$\pm$	16	\\
$\log g$ (cgs)	&	4.47	 $\pm$	0.06	\\
v$_\mathrm{turb}$ (km s$^{-1}$)	&	1.10	$\pm$	0.05	 \\
\hline
\unskip[Fe/H]	&	-0.25	$\pm$	0.04 	\\
\unskip[Na/H]	&	-0.21	$\pm$	0.08	 \\
\unskip[Mg/H]	&	-0.28	$\pm$	0.1	\\
 \unskip[Al/H]	&	-0.12	$\pm$	0.09	\\
 \unskip[Si/H]	&	-0.25	$\pm$	0.06	\\
 \unskip[Ca/H]	&	-0.20	 $\pm$	0.05	\\
 \unskip[Sc/H]	&	-0.21	$\pm$	0.05	\\
 \unskip[Ti/H]	&	-0.15	$\pm$	0.06	\\
 \unskip[V/H]	&	-0.14	$\pm$	0.09	\\
 \unskip[Cr/H]	&	-0.23	$\pm$	0.05	\\
 \unskip[Mn/H]	&	-0.33	$\pm$	0.08	\\
 \unskip[Co/H]	&	-0.20 	$\pm$	0.05	\\
 \unskip[Ni/H]	&	-0.24	$\pm$	0.06	\\
 \unskip[Y/H]	&	0	$\pm$	0.05	\\
 \unskip[Ba/H]	&	0.28	 $\pm$	0.05	\\
\hline                  
\end{tabular}

*For the fundamental parameters, the $\sigma$-value represents the intrinsic uncertainties computed following \citep{gonzalez}, while for chemical abundances $\sigma$ is the standard deviation around the mean abundance obtained from all the measured lines.
\end{table}

\subsection{Photometric parameters}

To determine the most precise set of photometric parameters of the system, all the 49 light curves were modelled with the version 34 of the JKTEBOP\footnote{http://www.astro.keele.ac.uk/jkt/codes/jktebop.html} code \citep{sou04}. This code uses a Levenberg-Marquardt optimisation algorithm to get the model that best fits a transit, and includes Monte Carlo and bootstrapping routines for error analysis.  For each transit we fitted the inclination ($i$), the sum of the fractional radii ($\Sigma=r_{\star}+r_\mathrm{P}$)\footnote{$r_{\star} =\frac{R_{\star}}{a}$ and $r_\mathrm{P} =\frac{R_{P}}{a}$ are the ratios of the absolute radii of the star and the exoplanet, respectively, to the semimajor axis ($a$).}, the ratio of the fractional radii ($k = r_{P}/r_{\star}$), the scale factor ($l_{0}$)\footnote{This parameter controls the flux level of the out-of-transit data-points in the light curves.}, and the mid-transit time ($T_\mathrm{0}$). This new version of JKTEBOP allows polynomial fitting (up to fifth order) and sine curves simultaneously to the modelling. Therefore, to remove the smooth trends in the light curves caused by differences between the spectral types of the comparison
and the exoplanet host-star, differential extinction, and stellar activity, we also fitted simultaneously the coefficients of a second order polynomial.  Both orbital period ($P$) and eccentricity ($e$) were kept as fixed parameters. For the photometric parameters ($i$, $\Sigma$, and $k$) and $P$ we adopted as initial values those obtained in \citet{ciceri}, while $e$  was assumed to be zero as it was determined in \citet{anderson} and $l_{0}$ equals to 1. All the light curves were modelled considering a quadratic limb-darkening (LD) law. The values of the linear and quadratic LD coefficients, $q_{1}$ and $q_{2}$ respectively, were computed by bilinearly interpolating $T_\mathrm{eff}$ and $\log g$ from the tables of \citet{claret} using the program JKTLD\footnote{http://www.astro.keele.ac.uk/jkt/codes/jktld.html}. However, these tabulations do not provide theoretical LD coefficients for several of the filters used to obtain the transits. In those cases, we adopted the tabulated values of filters with central wavelengths close to those in which the observations were made. Therefore, for the light curves observed in the $g'r'i'z'$ bands with GROND, we used the values of the $g'r'i'z'$ SLOAN filters. For those transits acquired with the Johnson $R$, the Gunn $r'$, and the Bessell $R$ filters, we used the values tabulated for the Cousin $R$ filter, and for the light curves obtained with no filter we used the average of the values corresponding to the Johnson $V$ and the Cousin $R$ bands. For the transit observed in the Gunn $g'$ band we used the SLOAN $g'$ filter. Finally, for the observation made in the Cousins $I+$Sloan $z'$ band, we adopted the values of the Sloan $z'$ filter.

To achieve the best-fitting model for each transit, we examined the results obtained considering: a) both LD coefficients as free parameters, b) the linear coefficient slightly perturbed\footnote{As indicated in \citet{sou12}, the coefficient was perturbed by $\pm$ 0.10 around the initial value in the error analysis simulations.} and the quadratic one freely varying, and c) both LD coefficients fixed at their initial values. In almost all the cases, option "a" gave unphysical results. Among the three possibilities, we choose the option that provided realistic parameters and the lowest value for the $\chi^{2}_{r}$. Photometric errors were multiplied by the square-root of the reduced chi-squared of the fit to get $\chi^{2}_{r}=1$. As a final step, we ran 10000 Monte Carlo (MC) iterations and a residual permutation algorithm which considers the presence of red noise in the photometric data. We adopted as the best-fitting parameters for each transit the median values of the algorithm (MC or residual permutation) that resulted in the largest error, while their errors are the asymmetric uncertainties $\sigma_{+}$ and $\sigma_{-}$, defined by a range of 68.3$\%$ values of the selected distribution. In Table 4 we list the parameters for all the light curves. 

We also evaluated the quality of each light curve, through the photometric noise rate (PNR), defined by \citet{fulton} as,

\begin{equation}
      PNR = \frac{\mathrm{RMS}}{\sqrt{ \Gamma}},
\end{equation}

\noindent where $\mathrm{RMS}$ is the standard deviation of the light curve residuals obtained by subtracting the JKTEBOP model from the photometric data, and $\Gamma$ is the median number of exposures per minute. The red noise level was also estimated through the $\beta$ parameter, which is defined by \citet{winn08} as $\beta=\frac{\sigma_{\mathrm{r}}}{\sigma_{\mathrm{N}}}$. Here, $\sigma_{\mathrm{r}}$ is determined by averaging the residuals into M bins of N points each and computing the standard deviation of the binned residuals, and $\sigma_{\mathrm{N}}$ represents the expected standard deviation, which in the absence of red noise is defined as,

\begin{equation*}
      \sigma_{\mathrm{N}} = \frac{\sigma_{\mathrm{1}}}{\sqrt{N}}\sqrt{\frac{M}{M-1}},
\end{equation*}

\noindent where $\sigma_{\mathrm{1}}$ is the standard deviation of the unbinned out-of-transit data. Since for WASP-46b the ingress/egress duration of the transit is $\sim$ 26 minutes, residuals were averaged in bins of between 16 and 36 minutes, and the median value was the adopted red noise factor. In Fig. 3 we show a plot of RMS as a function of the light curve bin size used to calculate the values of $\beta$ for the 12 new transits. Magenta, blue and green lines represent the measured standard deviations of the binned residuals ($\sigma_\mathrm{r}$) for the THG, EABA, and CASLEO data, respectively, and the black lines correspond to the expected standard deviations ($\sigma_\mathrm{N}$). This plot along with columns 8 and 9 of Table 4 show that our whole sample is composed of light curves of different quality and red noise level. 

Taking this into account, to avoid that final photometric parameters being affected by the results obtained from poor quality light curves, we estimated the best set of photometric parameters for the WASP-46 system considering only our most precise complete transits, i.e., light curves with PNR $\le$ 3 and $\beta$ $\le$ 1.25\footnote{These values of PNR and $\beta$ used to distinguish between high and poor quality transits, were arbitrarily determined from the available light curves to perform this study.}.  With this criterion we choose a total of 22 light curves\footnote{Although the light curves observed the nights of 2011, August 31 and 2012, July 27 have  PNR $\le$ 3 and $\beta$ $\le$ 1.25, we did not include them in the analysis because they present anomalies during the transit probably produced by the passage of the planet in front of starspots.} (indicated in Table 4 with asterisks) and computed the parameters of the system $i$, $\Sigma$, and $k$ as the weighted average of the values determined for each of the selected transits. Parameters uncertainties were calculated as the standard deviation of the sample relative to the number of data. From the 3$^{rd}$ law of \textit{Kepler} and assuming $M_\mathrm{P}$ $\ll$ $M_{\star}$ (where $M_{\star}$ is the stellar mass), we also estimated the mean stellar density using

\begin{equation}
      \rho_{\star} = \frac{4 \pi^{2}}{G P^{2}} \left(\frac{1}{r_{\star}} \right)^{3},
\end{equation}

\noindent where $G$ represents the gravitational constant and the value adopted for the orbital period is the one obtained with Eq. (5) of Section 4. The stellar-density uncertainty was computed through error's propagation. 
In Table 5 we present our results compared with the values previously obtained by \citet{anderson}, \citet{kjurk}, and \citet{ciceri}. Our parameters agree within the errors with those determined by \citet{anderson}, but are slightly different from the ones measured by \citet{ciceri} (except for the value of $k$ which is fully consistent), and those obtained by \citet{kjurk}. 

\subsection{Physical parameters}

Physical parameters were calculated using the JKTABSDIM\footnote{http://www.astro.keele.ac.uk/jkt/codes/jktabsdim.html} code \citep{sou09}, as explained in \citet{petrucci13, petrucci15}. Briefly, this procedure requires as input certain photometric and spectroscopic parameters with their errors. In particular, the best value for the velocity
amplitude of the planet is determined by linearly interpolating this parameter between three different stellar models:  Y$^2$ \citep{dem04}, Padova \citep{girardi}, and Teramo \citep{pie04}. For each stellar model, we considered several isochrones comprising the lifetime of the star in the main-sequence. In particular, we used isochrones in the ranges 1 Myr -- 20 Gyr, 63 Myr -- 18 Gyr, and 30 Myr -- 16 Gyr for the Y$^2$, Padova, and Teramo models, respectively. Then, we calculated the values of $M_{\star}$, $R_{\star}$, log g$_{\star}$, $M_\mathrm{P}$, $R_\mathrm{P}$, $a$, and age with their respective errors.

Planetary surface gravity ($g_\mathrm{P}$), modified equilibrium temperature ($T'_{eq}$), and Safronov number ($\Theta$) were determined independently on the stellar models, using Eq. (4) of \citet{sou07} and Eqs. (5) and (6) of \citet{sou10}. The modified equilibrium temperature is similar to the equilibrium temperature (i.e. the temperature that would have a planet if it were supposed as a black body heated only by its parent star) when the Bond albedo, A, is considered equal to $1 - 4\mathrm{F}$, where F is a heat redistribution factor; while the Safronov number is an indicator of the efficiency with which a planet scatters other bodies \citep{fressin}. Parameters uncertainties were calculated from the propagation of errors. In Table 6 we present our results along with the values previously obtained by \citet{anderson} and \citet{ciceri}. It can be seen that, in general, our parameters are in good agreement with those computed in previous works. However, in the particular case of the planetary radius, our estimation is similar within the error to the value presented in the discovery paper \citep{anderson} but slightly larger than the result obtained by \citet{ciceri}. Finally, although the density calculated in this work for WASP-46b agrees within errors with the ones previously determined, our value indicates that the planet is more dense than  pointed out by \citet{anderson} but not as much as claimed by \citet{ciceri}.

\begin{figure}
   \centering
   \includegraphics[width=.5\textwidth]{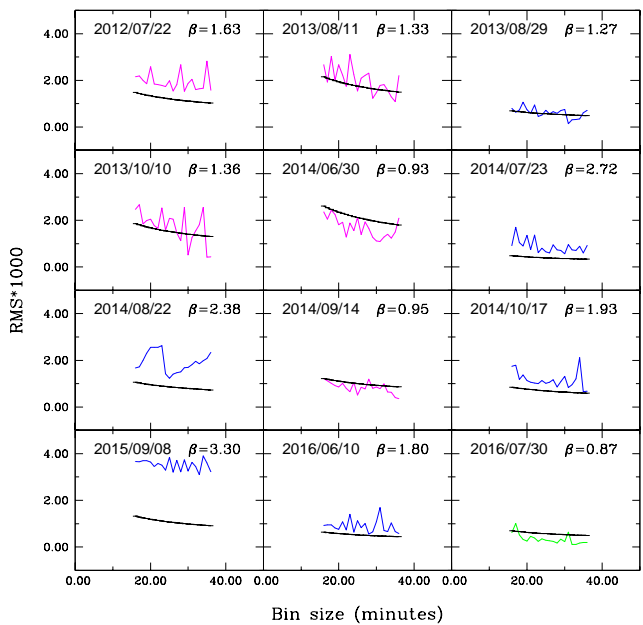}
      \caption{RMS vs light curve bin size used to calculate the values of red noise ($\beta$) for the 12 new transits presented in this work. Magenta, blue and green lines represent the measured standard deviations of the binned residuals ($\sigma_\mathrm{r}$) for the THG, EABA, and CASLEO data, respectively, and black lines correspond to the expected standard deviations ($\sigma_\mathrm{N}$). For each transit, the observation date and the computed red noise are also pointed out.}
         \label{FigVibStab}
   \end{figure}

\begin{table*}
{\scriptsize
\caption{Photometric parameters and quality factors determined for the 49 light curves analysed in this work}             
\label{table:4}      
\centering          
\begin{tabular}{lccccccccccc}     
\hline\hline       
Date	&	Epoch	&	$i$ 	&	$k$	& $\Sigma$ &	 $r_{\star}$ &	$r_\mathrm{P}$ & PNR & $\beta$ & Filter & Complete? & Reference \\
 &	&	($^{\circ}$) 	&		&  &	  &	 & (mmag) &  & & &  \\
\hline                    
Jul 19 2010$^{*}$	&	3	&	81.94	   $^{+	0.74	}_{-	0.79	}$ &	0.14116	   $^{+	0.00363	}_{-	0.00482	}$ &	0.20839	   $^{+	0.00995	}_{-	0.00919	}$ &	0.1824	   $^{+	0.0090	}_{-	0.0080	}$ &	0.02574	   $^{+	0.00108	}_{-	0.00124	}$ &	1.7506	&	0.5839	&	Cousins I+Sloan $z'$	&	Yes &	1 \\
Sep 10 2010$^{*}$	&	40	&	81.50	   $^{+	0.84	}_{-	0.35	}$ &	0.14047	   $^{+	0.00239	}_{-	0.00298	}$ &	0.21304	   $^{+	0.00568	}_{-	0.01244	}$ &	0.1869	   $^{+	0.0045	}_{-	0.0109	}$ &	0.02616	   $^{+	0.00089	}_{-	0.00162	}$ &	0.7430	&	0.9474	&	Gunn $r'$	&	Yes	& 1 \\
Jun 09 2011$^{*}$	&	231	&	83.02	   $^{+	1.10	}_{-	1.54	}$ &	0.13307	   $^{+	0.00465	}_{-	0.00845	}$ &	0.19628	   $^{+	0.01685	}_{-	0.01364	}$ &	0.1730	   $^{+	0.0151	}_{-	0.0114	}$ &	0.02322	   $^{+	0.00182	}_{-	0.00223	}$ &	0.8988	&	1.0153	&	Gunn $r'$	&	Yes	& 2\\
Jul 14 2011	&	255	&	83.06	   $^{+	2.76	}_{-	1.59	}$ &	0.15276	   $^{+	0.01017	}_{-	0.01084	}$ &	0.19089	   $^{+	0.02626	}_{-	0.03489	}$ &	0.1653	   $^{+	0.0215	}_{-	0.0289	}$ &	0.02553	   $^{+	0.00449	}_{-	0.00598	}$ &	11.4244	&	0.7333	&	clear	&	Yes & 3 \\
Aug 31 2011	&	289	&	84.45	   $^{+	1.13	}_{-	0.85	}$ &	0.13558	   $^{+	0.00407	}_{-	0.00675	}$ &	0.17825	   $^{+	0.01293	}_{-	0.00884	}$ &	0.1569	   $^{+	0.0123	}_{-	0.0076	}$ &	0.02116	   $^{+	0.00169	}_{-	0.00200	}$ &	0.4862	&	1.2194	&	Gunn $r'$	&	Yes & 2	\\
Oct 23 2011$^{*}$	&	326	&	81.95	   $^{+	0.37	}_{-	0.27	}$ &	0.14146	   $^{+	0.00151	}_{-	0.00261	}$ &	0.20532	   $^{+	0.00280	}_{-	0.00432	}$ &	0.1797	   $^{+	0.0028	}_{-	0.0037	}$ &	0.02542	   $^{+	0.00038	}_{-	0.00060	}$ &	0.2023	&	1.2283	&	Gunn $g'$	&	Yes	& 2\\
Jun 30 2012	&	501	&	84.69	   $^{+	5.15	}_{-	3.78	}$ &	0.13951	   $^{+	0.01043	}_{-	0.01011	}$ &	0.17478	   $^{+	0.05352	}_{-	0.03328	}$ &	0.1532	   $^{+	0.0445	}_{-	0.0281	}$ &	0.02168	   $^{+	0.00806	}_{-	0.00549	}$ &	3.3677	&	1.3711	&	clear	&	Yes	& 4\\
Jul 02 2012$^{*}$	&	503	&	82.43	   $^{+	0.32	}_{-	0.39	}$ &	0.14327	   $^{+	0.00144	}_{-	0.00222	}$ &	0.20130	   $^{+	0.00454	}_{-	0.00414	}$ &	0.1760	   $^{+	0.0042	}_{-	0.0034	}$ &	0.02526	   $^{+	0.00082	}_{-	0.00076	}$ &	0.3816	&	1.0983	&	Sloan $g'$	&	Yes & 2	\\
Jul 02 2012$^{*}$	&	503	&	82.82	   $^{+	0.42	}_{-	0.51	}$ &	0.14030	   $^{+	0.00148	}_{-	0.00119	}$ &	0.19527	   $^{+	0.00637	}_{-	0.00449	}$ &	0.1712	   $^{+	0.0056	}_{-	0.0038	}$ &	0.02415	   $^{+	0.00081	}_{-	0.00078	}$ &	0.3868	&	1.2257	&	Sloan $i'$	&	Yes	& 2 \\ 
Jul 02 2012$^{*}$	&	503	&	82.14	   $^{+	0.45	}_{-	0.37	}$ &	0.14321	   $^{+	0.00166	}_{-	0.00238	}$ &	0.20570	   $^{+	0.00412	}_{-	0.00521	}$ &	0.1797	   $^{+	0.0039	}_{-	0.0050	}$ &	0.02576	   $^{+	0.00062	}_{-	0.00067	}$ &	0.4301	&	1.0492	&	Sloan $r'$	&	Yes	& 2 \\
Jul 02 2012	&	503	&	84.17	   $^{+	0.58	}_{-	0.73	}$ &	0.13714	   $^{+	0.00285	}_{-	0.00252	}$ &	0.18397	   $^{+	0.00772	}_{-	0.00681	}$ &	0.1617	   $^{+	0.0065	}_{-	0.0056	}$ &	0.02228	   $^{+	0.00131	}_{-	0.00120	}$ &	0.5048	&	1.3005	&	Sloan $z'$	&	Yes & 2	\\
Jul 21 2012$^{*}$	&	516	&	81.76	   $^{+	1.69	}_{-	1.11	}$ &	0.14006	   $^{+	0.00473	}_{-	0.01475	}$ &	0.21933	   $^{+	0.01414	}_{-	0.02077	}$ &	0.1924	   $^{+	0.0128	}_{-	0.0170	}$ &	0.02702	   $^{+	0.00209	}_{-	0.00394	}$ &	1.8542	&	0.8195	&	$R$	&	Yes & 5 \\
Jul 22 2012	&	517	&	83.81	   $^{+	2.35	}_{-	2.90	}$ &	0.14653	   $^{+	0.01064	}_{-	0.00716	}$ &	0.18646	   $^{+	0.04319	}_{-	0.02438	}$ &	0.1625	   $^{+	0.0365	}_{-	0.0208	}$ &	0.02383	   $^{+	0.00794	}_{-	0.00397	}$ &	14.4072	&	1.6334	&	clear	&	Yes & 6	\\
Jul 26 2012	&	519	&	82.68	   $^{+	0.92	}_{-	0.62	}$ &	0.14002	   $^{+	0.00281	}_{-	0.00541	}$ &	0.19467	   $^{+	0.00715	}_{-	0.01027	}$ &	0.1705	   $^{+	0.0065	}_{-	0.0088	}$ &	0.02404	   $^{+	0.00094	}_{-	0.00186	}$ &	1.3726	&	1.2339	&	clear	&	Yes	& 7 \\
Sep 23 2012$^{*}$	&	561	&	83.52	   $^{+	0.52	}_{-	0.96	}$ &	0.14312	   $^{+	0.00393	}_{-	0.00533	}$ &	0.18859	   $^{+	0.01133	}_{-	0.00830	}$ &	0.1652	   $^{+	0.0091	}_{-	0.0073	}$ &	0.02358	   $^{+	0.00200	}_{-	0.00181	}$ &	0.9753	&	0.9333	&	Bessell $R$	&	Yes & 2	\\
Oct 16 2012$^{*}$	&	577	&	83.40	   $^{+	0.69	}_{-	0.57	}$ &	0.13681	   $^{+	0.00333	}_{-	0.00541	}$ &	0.19302	   $^{+	0.00495	}_{-	0.00704	}$ &	0.1701	   $^{+	0.0039	}_{-	0.0061	}$ &	0.02321	   $^{+	0.00105	}_{-	0.00145	}$ &	0.5359	&	0.7216	&	Sloan $g'$	&	Yes	& 2 \\
Oct 16 2012$^{*}$	&	577	&	82.92	   $^{+	0.36	}_{-	0.43	}$ &	0.14130	   $^{+	0.00153	}_{-	0.00209	}$ &	0.19481	   $^{+	0.00505	}_{-	0.00432	}$ &	0.1708	   $^{+	0.0043	}_{-	0.0039	}$ &	0.02423	   $^{+	0.00062	}_{-	0.00069	}$ &	0.5234	&	0.9718	&	Sloan $z'$	&	Yes & 2	\\
Oct 27 2012	&	584	&	84.88	   $^{+	2.51	}_{-	1.48	}$ &	0.13970	   $^{+	0.00722	}_{-	0.00704	}$ &	0.17013	   $^{+	0.02088	}_{-	0.02474	}$ &	0.1494	   $^{+	0.0175	}_{-	0.0212	}$ &	0.02114	   $^{+	0.00367	}_{-	0.00373	}$ &	3.0259	&	0.9244	&	clear	&	Yes	& 8\\
Apr 24 2013$^{*}$	&	710	&	83.49	   $^{+	0.88	}_{-	0.71	}$ &	0.13021	   $^{+	0.01410	}_{-	0.00451	}$ &	0.18876	   $^{+	0.01784	}_{-	0.00967	}$ &	0.1663	   $^{+	0.0170	}_{-	0.0094	}$ &	0.02263	   $^{+	0.00166	}_{-	0.00145	}$ &	0.3543	&	0.8296	&	Sloan $g'$	&	Yes & 2	\\
Apr 24 2013	&	710	&	82.89	   $^{+	0.57	}_{-	0.32	}$ &	0.14176	   $^{+	0.00206	}_{-	0.00291	}$ &	0.19198	   $^{+	0.00498	}_{-	0.00763	}$ &	0.1681	   $^{+	0.0048	}_{-	0.0070	}$ &	0.02388	   $^{+	0.00064	}_{-	0.00111	}$ &	0.5557	&	1.2993	&	Sloan $i'$	&	Yes & 2	\\
Apr 24 2013	&	710	&	81.74	   $^{+	0.44	}_{-	0.36	}$ &	0.14644	   $^{+	0.00223	}_{-	0.00338	}$ &	0.21186	   $^{+	0.00441	}_{-	0.00660	}$ &	0.1848	   $^{+	0.0041	}_{-	0.0062	}$ &	0.02691	   $^{+	0.00077	}_{-	0.00088	}$ &	0.3502	&	1.4976	&	Sloan $r'$	&	Yes	& 2\\
Apr 24 2013	&	710	&	85.12	   $^{+	0.84	}_{-	0.82	}$ &	0.15740	   $^{+	0.00682	}_{-	0.00542	}$ &	0.16804	   $^{+	0.00715	}_{-	0.00816	}$ &	0.1452	   $^{+	0.0070	}_{-	0.0077	}$ &	0.02314	   $^{+	0.00075	}_{-	0.00132	}$ &	0.4398	&	1.3203	&	Sloan $z'$	&	Yes & 2	\\
Jun 16 2013$^{*}$	&	747	&	82.45	   $^{+	0.68	}_{-	0.57	}$ &	0.14001	   $^{+	0.00153	}_{-	0.00536	}$ &	0.20049	   $^{+	0.00767	}_{-	0.01008	}$ &	0.1758	   $^{+	0.0076	}_{-	0.0086	}$ &	0.02463	   $^{+	0.00112	}_{-	0.00123	}$ &	0.5489	&	0.8654	&	Sloan $g'$	&	Yes	& 2 \\
Jun 16 2013$^{*}$	&	747	&	83.67	   $^{+	0.65	}_{-	0.72	}$ &	0.13813	   $^{+	0.00248	}_{-	0.00241	}$ &	0.18213	   $^{+	0.00902	}_{-	0.00996	}$ &	0.1599	   $^{+	0.0079	}_{-	0.0085	}$ &	0.02215	   $^{+	0.00122	}_{-	0.00154	}$ &	0.7442	&	0.8504	&	Sloan $i'$	&	Yes & 2	\\
Jun 16 2013$^{*}$	&	747	&	81.89	   $^{+	0.43	}_{-	0.59	}$ &	0.13653	   $^{+	0.00332	}_{-	0.00293	}$ &	0.20542	   $^{+	0.00747	}_{-	0.00673	}$ &	0.1806	   $^{+	0.0070	}_{-	0.0065	}$ &	0.02473	   $^{+	0.00113	}_{-	0.00111	}$ &	0.7624	&	1.0033	&	Sloan $r'$	&	Yes	& 2\\
Jun 16 2013	&	747	&	83.42	   $^{+	2.80	}_{-	1.68	}$ &	0.13587	   $^{+	0.00842	}_{-	0.01751	}$ &	0.19649	   $^{+	0.01529	}_{-	0.02736	}$ &	0.1723	   $^{+	0.0141	}_{-	0.0239	}$ &	0.02338	   $^{+	0.00268	}_{-	0.00445	}$ &	1.0829	&	1.6236	&	Sloan $z'$	&	Yes & 2	\\
Aug 05 2013$^{*}$	&	782	&	82.84	   $^{+	0.55	}_{-	0.73	}$ &	0.13305	   $^{+	0.01271	}_{-	0.00589	}$ &	0.20166	   $^{+	0.01508	}_{-	0.00940	}$ &	0.1776	   $^{+	0.0146	}_{-	0.0094	}$ &	0.02416	   $^{+	0.00118	}_{-	0.00102	}$ &	0.6023	&	0.8213	&	Bessell $R$	&	Yes & 2	\\
Aug 05 2013$^{*}$	&	782	&	82.30	   $^{+	1.39	}_{-	1.19	}$ &	0.14998	   $^{+	0.00682	}_{-	0.00671	}$ &	0.20718	   $^{+	0.02177	}_{-	0.01589	}$ &	0.1801	   $^{+	0.0175	}_{-	0.0131	}$ &	0.02701	   $^{+	0.00368	}_{-	0.00240	}$ &	2.9548	&	1.0688	&	clear	&	Yes & 4	\\
Aug 11 2013	&	786	&	87.08	   $^{+	2.82	}_{-	5.38	}$ &	0.13725	   $^{+	0.01154	}_{-	0.02501	}$ &	0.16341	   $^{+	0.04504	}_{-	0.02108	}$ &	0.1440	   $^{+	0.0403	}_{-	0.0192	}$ &	0.01918	   $^{+	0.00528	}_{-	0.00244	}$ &	7.0480	&	1.3326	&	Jhonson $R$ 	&	Yes	& 6\\
Aug 16 2013 &	789	&	84.52	   $^{+	3.77	}_{-	2.52	}$ &	0.13968	   $^{+	0.00637	}_{-	0.01932	}$ &	0.17521	   $^{+	0.03155	}_{-	0.02555	}$ &	0.1538	   $^{+	0.0284	}_{-	0.0219	}$ &	0.02152	   $^{+	0.00259	}_{-	0.00517	}$ &	4.9027	&	0.8861	&	clear	&	Yes & 4	\\
Aug 28 2013$^{*}$	&	798	&	82.66	   $^{+	0.59	}_{-	0.49	}$ &	0.14033	   $^{+	0.00149	}_{-	0.00215	}$ &	0.19641	   $^{+	0.00554	}_{-	0.00619	}$ &	0.1722	   $^{+	0.0051	}_{-	0.0053	}$ &	0.02425	   $^{+	0.00065	}_{-	0.00105	}$ &	0.6097	&	1.1871	&	Bessell $R$	&	Yes & 2	\\
Aug 28 2013	&	798	&	86.33	   $^{+	3.55	}_{-	2.41	}$ &	0.12691	   $^{+	0.00492	}_{-	0.00378	}$ &	0.15459	   $^{+	0.02522	}_{-	0.01578	}$ &	0.1374	   $^{+	0.0216	}_{-	0.0138	}$ &	0.01722	   $^{+	0.00236	}_{-	0.00208	}$ &	2.8194	&	1.2699	&	Jhonson $R$ 	&	Yes & 9	\\
Oct 10 2013 	&	828	&	81.50	   $^{+	4.71	}_{-	2.09	}$ &	0.13207	   $^{+	0.01253	}_{-	0.01339	}$ &	0.21674	   $^{+	0.03140	}_{-	0.06531	}$ &	0.1910	   $^{+	0.0255	}_{-	0.0561	}$ &	0.02523	   $^{+	0.00603	}_{-	0.00945	}$ &	7.1662	&	1.3588	&	Jhonson $V$ 	&	Yes & 6	\\
Oct 24 2013 	&	837	&	84.12	   $^{+	1.30	}_{-	1.44	}$ &	0.13709	   $^{+	0.00443	}_{-	0.00834	}$ &	0.18078	   $^{+	0.01809	}_{-	0.01728	}$ &	0.1585	   $^{+	0.0162	}_{-	0.0143	}$ &	0.02163	   $^{+	0.00288	}_{-	0.00280	}$ &	3.1192	&	1.1956	&	clear	&	Yes & 10	\\
Nov 13 2013	&	851	&	83.53	   $^{+	1.17	}_{-	1.20	}$ &	0.13452	   $^{+	0.00583	}_{-	0.00394	}$ &	0.18749	   $^{+	0.01646	}_{-	0.01749	}$ &	0.1642	   $^{+	0.0151	}_{-	0.0139	}$ &	0.02214	   $^{+	0.00284	}_{-	0.00241	}$ &	3.0851	&	1.6816	&	clear	&	Yes & 10	\\
Jun 30 2014	&	1012	&	86.53	   $^{+	3.38	}_{-	4.84	}$ &	0.15161	   $^{+	0.00666	}_{-	0.02164	}$ &	0.15054	   $^{+	0.05371	}_{-	0.01725	}$ &	0.1310	   $^{+	0.0494	}_{-	0.0156	}$ &	0.01959	   $^{+	0.00437	}_{-	0.00219	}$ &	9.7489	&	0.9315	&	Jhonson $R$ 	&	Yes	& 6\\
Jul 23 2014	&	1028	&	82.96	   $^{+	0.89	}_{-	0.83	}$ &	0.14754	   $^{+	0.00417	}_{-	0.00567	}$ &	0.19718	   $^{+	0.01067	}_{-	0.01389	}$ &	0.1716	   $^{+	0.0096	}_{-	0.0118	}$ &	0.02536	   $^{+	0.00181	}_{-	0.00219	}$ &	6.7399	&	2.7261	&	Jhonson $R$ 	&	Yes & 9	\\
Aug 13 2014	&	1042	&	79.55	   $^{+	1.79	}_{-	3.38	}$ &	0.17084	   $^{+	0.21292	}_{-	0.01595	}$ &	0.24401	   $^{+	0.04051	}_{-	0.02944	}$ &	0.1995	   $^{+	0.0166	}_{-	0.0174	}$ &	0.03648	   $^{+	0.04056	}_{-	0.00755	}$ &	3.4746	&	1.3159	&	clear	&	Yes & 11	\\
Aug 16 2014$^{*}$ &	1044	&	82.30	   $^{+	1.62	}_{-	1.23	}$ &	0.14171	   $^{+	0.00907	}_{-	0.01431	}$ &	0.20461	   $^{+	0.02200	}_{-	0.02110	}$ &	0.1791	   $^{+	0.0197	}_{-	0.0178	}$ &	0.02530	   $^{+	0.00328	}_{-	0.00348	}$ &	2.6317	&	0.5278	&	clear	&	Yes & 11	\\
Aug 22 2014	&	1049	&	82.28	   $^{+	1.84	}_{-	2.23	}$ &	0.14720	   $^{+	0.00942	}_{-	0.01137	}$ &	0.20243	   $^{+	0.03689	}_{-	0.02423	}$ &	0.1764	   $^{+	0.0310	}_{-	0.0208	}$ &	0.02568	   $^{+	0.00643	}_{-	0.00369	}$ &	2.7173	&	2.3797	&	Jhonson $R$ 	&	No & 9	\\
Sep  14 2014	&	1065	&	88.93	   $^{+	0.97	}_{-	7.25	}$ &	0.13904	   $^{+	0.01337	}_{-	0.01112	}$ &	0.17971	   $^{+	0.02003	}_{-	0.01447	}$ &	0.1575	   $^{+	0.0191	}_{-	0.0113	}$ &	0.02211	   $^{+	0.00275	}_{-	0.00358	}$ &	4.3655	&	0.9495	&	Jhonson $R$ 	&	Yes & 6	\\
Oct 12 2014	&	1084	&	84.11	   $^{+	2.57	}_{-	1.34	}$ &	0.16448	   $^{+	0.00677	}_{-	0.00962	}$ &	0.17812	   $^{+	0.01959	}_{-	0.03041	}$ &	0.1525	   $^{+	0.0170	}_{-	0.0245	}$ &	0.02520	   $^{+	0.00336	}_{-	0.00557	}$ &	4.0209	&	0.9035	&	clear	&	Yes & 4	\\
Oct 17 2014	&	1088	&	81.73	   $^{+	1.34	}_{-	1.20	}$ &	0.15391	   $^{+	0.00753	}_{-	0.01176	}$ &	0.22223	   $^{+	0.02310	}_{-	0.02208	}$ &	0.1926	   $^{+	0.0197	}_{-	0.0189	}$ &	0.02943	   $^{+	0.00351	}_{-	0.00343	}$ &	7.2769	&	1.9345	&	Jhonson $R$ 	&	Yes & 9	\\
Sep 08 2015	&	1316	&	85.49	   $^{+	4.34	}_{-	2.68	}$ &	0.13164	   $^{+	0.02183	}_{-	0.01697	}$ &	0.18494	   $^{+	0.05226	}_{-	0.03986	}$ &	0.1623	   $^{+	0.0454	}_{-	0.0345	}$ &	0.02137	   $^{+	0.00691	}_{-	0.00563	}$ &	6.4132	&	3.3073	&	Jhonson $R$ 	&	Yes & 9	\\
Sep 28 2015 &	1330	&	82.10	   $^{+	1.66	}_{-	2.57	}$ &	0.23812	   $^{+	0.06121	}_{-	0.03058	}$ &	0.22861	   $^{+	0.03427	}_{-	0.03046	}$ &	0.1858	   $^{+	0.0146	}_{-	0.0248	}$ &	0.04346	   $^{+	0.01604	}_{-	0.00894	}$ &	7.5577	&	0.9996	&	clear	&	Yes & 11	\\
Jun 10 2016	&	1509	&	82.25	   $^{+	0.68	}_{-	0.57	}$ &	 0.14693	   $^{+	0.00474	}_{-	0.00312	}$ &	0.20436	   $^{+	0.00906	}_{-	0.0101	}$ &	 0.17806	   $^{+	0.00754	}_{-	0.00845	}$ &	 0.02622	   $^{+	0.00171	}_{-	0.00156	}$ &	 2.6218	&	1.8010	&	Jhonson R	&	Yes & 9	\\
Jul 22 2016$^{*}$	&	1539 	&	83.78	   $^{+	0.96	}_{-	1.44	}$ &	 0.13441	   $^{+	0.00275	}_{-	0.00501	}$ &	 0.18119	   $^{+	0.01629	}_{-	0.01303	}$ &	 0.15961	   $^{+	0.01369	}_{-	0.01101	}$ &	0.02149	   $^{+	0.00221	}_{-	0.00178	}$ &	 2.7467	&	0.9531	&	clear	&	Yes & 12	\\
Jul 30 2016$^{*}$	&	1544	 &	81.6	   $^{+	1.75	}_{-	1.29	}$ &	 0.14653	   $^{+	0.00379	}_{-	0.00732	}$ &	 0.21995	   $^{+	0.01387	}_{-	0.02119	}$ &	 0.19175	   $^{+	0.01266	}_{-	0.0179	}$ &	 0.02783	   $^{+	0.002	}_{-	0.00321	}$ &	 2.8966	&	0.8727	&	Jhonson R	&	Yes & 13	\\
Aug 08 2016$^{*}$	&	1551 	&	82.73   $^{+	0.41	}_{-	0.43	}$ &	 0.14314	   $^{+	0.00383	}_{-	0.00293	}$ &	0.19951	   $^{+	0.00657	}_{-	0.00667	}$ &	 0.17426	   $^{+	0.00582	}_{-	0.00567	}$ &	0.02489	   $^{+	0.00127	}_{-	0.001	}$ &	 2.1361	&	0.8638	&	clear	&	Yes & 14	\\

\hline      
\end{tabular}}

Column 2: Values of the parameter Epoch, which represents the number of transits since the minimum reference time shown in Eq. (4). Columns 3-7: Values of the derived photometric parameters orbital inclination ($i$), ratio of the fractional radii ($k$), sum of the fractional radii ($\Sigma$), ratio of the absolute stellar radius to the semimajor axis ($r_{\star}$), and ratio of the absolute planetary radius to the semimajor axis ($r_\mathrm{P}$), and their errors. Column 8: Photon noise rate. Column 9: Median value for the red noise. 

References: (1) \citet{anderson}; (2) \citet{ciceri}; (3) Curtis I. (ETD); (4) Evans P. (ETD); (5) Sauer T. (ETD); (6) This work (THG); (7) Schneiter M, Colazo C. (ETD); (8) Colazo C. A. (ETD); (9) This work (EABA); (10) Schneiter M., Villarreal C., Colazo C. (ETD); (11) Ma\v{s}ek M., Ho\v{n}kov\'a K., Jury\v{s}ek J. (ETD); (12) Villarreal C., Qui\~nones C. (ETD); (13) This work (CASLEO); (14) Qui\~nones C., Melia R., Colazo C. (ETD).

Note: Asterisks indicate the transits used to calculate the final values of $i$, $k$, $\Sigma$, $r_{\star}$, and $r_\mathrm{P}$.
 
\end{table*}

\begin{table*}
\caption{Photometric parameters derived in this work along with the values previously determined by \citet{anderson}, \citet{kjurk}, and \citet{ciceri}}             
\label{table:5}      
\centering          
\begin{tabular}{lcccc}     
\hline\hline       
Parameter & This work	&	\citet{ciceri} & \citet{kjurk}	& \citet{anderson}\\
\hline                    
Orbital inclination, $i$ ($^{\circ}$)	&	82.53 $\pm$ 0.13 	&	82.80 $\pm$ 0.17	& 82.015 $\pm$ 0.005 	& 82.63 $\pm$ 0.38 \\
Ratio of fractional radii, $k$	&	0.14074	$\pm$ 0.00068	&	0.14075 $\pm$ 0.00035	& -- &	0.1468 $\pm$ 0.0017 \\
Sum of fractional radii, $r_{\star}$+$r_\mathrm{P}$	&	0.1999 $\pm$ 	0.0016	&	0.1950 $\pm$ 0.0013	& -- &	0.1992 $\pm$ 0.0059 \\
Stellar fractional radius, $r_{\star}$	&	0.1750	$\pm$ 0.0014 	&	0.1709 $\pm$ 0.0011	& 0.179 $\pm$ 0.001 &	 0.1742 $\pm$ 0.0057 \\
Planetary fractional radius, $r_\mathrm{P}$	&	0.02474	$\pm$ 0.00024	&	0.02403 $\pm$ 0.00021	& 0.02725 $\pm$ 0.00005	& 0.0250 $\pm$ 0.0010 \\
Stellar density, $\rho_{\star}$ ($\rho_{\odot}$) 	&	1.220	$\pm$ 0.031 	&	1.310 $\pm$ 0.025	& --	& 1.24 $\pm$ 0.10 \\
\hline                  
\end{tabular}

Note: Photometric errors of \citet{kjurk} might be underestimated since they are just the formal values measured by the code used to fit the transits.
\end{table*}

\begin{table*}
\caption{Physical parameters derived in this work along with the values previously determined by \citet{anderson} and \citet{ciceri}}             
\label{table:6}      
\centering          
\begin{tabular}{lccc}     
\hline\hline       
Parameter & This work	&	\citet{ciceri} &	 \citet{anderson}\\
\hline                    
Stellar mass, $M_{\star}$ ($M_{\odot}$)	&	0.907 $\pm$ 0.033 &	0.828 $\pm$ 0.067 $\pm$ 0.036	&	0.956 $\pm$ 0.034  \\
Stellar radius, $R_{\star}$ ($R_{\odot}$)	&	0.905 $\pm$ 0.013 &	0.858 $\pm$ 0.024 $\pm$ 0.013 	&	0.917 $\pm$ 0.028  \\
Planetary mass, $M_\mathrm{P}$ ($M_\mathrm{J}$)	&	2.031 $\pm$	0.072  	&	1.91 $\pm$ 0.11 $\pm$ 0.06 	&	2.101 $\pm$ 0.073 \\
Planetary radius, $R_\mathrm{P}$ ($R_\mathrm{J}$)	&	1.244 $\pm$	0.019	&	1.174 $\pm$ 0.033 $\pm$ 0.017 	&	1.310 $\pm$ 0.051  \\
Planetary surface gravity, $g_\mathrm{P}$ (m s$^{-2}$)	&	32.4 $\pm$	3.0 	&	34.3 $\pm$ 1.1 	&	28.0$^{+2.2}_{-2.0}$ \\
Planetary density, $\rho_{P}$ ($\rho_{J}$)	&	1.054 $\pm$ 0.062 	&	1.103 $\pm$ 0.050 $\pm$ 0.016 	&	0.94 $\pm$ 0.11  \\
Planetary modified equilibrium temperature, $T'_{eq}$ (K)	&	1704 $\pm$ 18	&	1636 $\pm$ 44 	&	1654 $\pm$ 50 \\
Safronov number, $\Theta$	&	0.0863 $\pm$ 0.0045 	&	0.0916 $\pm$ 0.0035 $\pm$ 0.0014 	&	-- \\
Semimajor axis, $a$ (AU)	&	0.02407 $\pm$ 0.00029 	&	0.02335 $\pm$ 0.00063 $\pm$ 0.00034 	&	0.02448 $\pm$ 0.00028 \\
Age (Gyr)	&	3.6	$\pm$ 1.9	&	9.6$^{+3.4 +1.4}_{-4.2 -3.5}$	&	 0.9-1.4$^{a}$ Gyr \\
\hline                  
\end{tabular}

$^{a}$: This is the value obtained by \citet{anderson} for the gyrochronological age.

Note: The modified equilibrium temperature is similar to the equilibrium temperature (i.e. the temperature that would have a planet if it were supposed as a black body heated only by its parent star) when the Bond albedo, A, is considered equal to $1 - 4\mathrm{F}$, where F is a heat redistribution factor; while the Safronov number is an indicator of the efficiency with which a planet scatters other bodies \citep{fressin}.
\end{table*}

\section{ANALYSIS OF TRANSIT TIMING VARIATIONS}

We used the mathematical transformations described in \citet{eastman} to convert measured times to $BJD_\mathrm{TDB}$. To carry out a fully homogeneous analysis of transit timing variations, mid-transit times were determined by fitting all the 49 light curves with the JKTEBOP code. Since $T_\mathrm{0}$ is not correlated with the photometric parameters, each individual light curve was modelled by assuming $T_\mathrm{0}$, $l_{0}$, and the coefficients of the polynomial to fit the out-of-transit data-points as the only free variables. As initial values for $i$, $k$, and $\Sigma$ we assumed those determined in Section 3.2. We ran 10000 MC iterations and a residual permutation algorithm. For the mid-transit times, we finally adopted the mean values given by the best JKTEBOP fit to each light curve, and the errors were assumed as the asymmetric uncertainties $\sigma_{+}$ and $\sigma_{-}$  of the algorithm (MC or residual permutation) that resulted in the largest error. These results are shown in Table 7. In our sample, we have independent observations of the same transit for two different epochs. Although the values of $T_\mathrm{0}$ do not agree with each other, even considering errors, it is important to mention that, in both cases, one of the transits has much higher quality than the other (PNR $\sim$ 0.6 compared to PNR $\sim$ 3). As we show in the next paragraphs, larger values of PNR imply less accurate values of $T_\mathrm{0}$, which should explain the discrepancy. In the following analysis, we treated each measurement separately.

\begin{table}
\caption{Mid-transit times calculated from the procedure explained in Section 4}             
\label{table:7}      
\centering                          
\begin{tabular}{c c c}        
\hline\hline                 
Epoch & $T_{\mathrm{0}}\ (BJD_{\mathrm{TDB}})$ & $e_{\mathrm{T_{\mathrm{0}}}}$ \\  
\hline                        
3	& 	2455396.607854	 & 	0.000618	 \\
40	& 	2455449.530824	 & 	0.000265	 \\
231	& 	2455722.731779	 & 	0.000229	 \\
255	& 	2455757.061951	 & 	0.000941	 \\
289$^{*}$	& 	2455805.693185	 & 	0.000205	 \\
326	& 	2455858.618330	 & 	0.000089	 \\
501	& 	2456108.927705	 & 	0.000941	 \\
503	& 	2456111.794218	 & 	0.000159	 \\
503	& 	2456111.794128	 & 	0.000120	 \\
503	& 	2456111.794240	 & 	0.000150	 \\
503$^{*}$	& 2456111.794547	 & 	0.000167	 \\
516	& 	2456130.388946	 & 	0.000415	 \\
517	& 	2456131.814561	 & 	0.001116	 \\
519$^{*}$	&	2456134.676108	 & 	0.000289	 \\
561	& 	2456194.759161	 & 	0.000270	 \\
577	& 	2456217.641274	 & 	0.000147	 \\
577	& 	2456217.641561	 & 	0.000132	 \\
584	& 	2456227.655743	 & 	0.000604	 \\
710	& 	2456407.880958	 & 	0.000154	 \\
710	& 	2456407.880849	 & 	0.000176	 \\
710	& 	2456407.881483	 & 	0.000282	 \\
710	& 	2456407.881594	 & 	0.000426	 \\
747	& 	2456460.805257	 & 	0.000173	 \\
747	& 	2456460.805193	 & 	0.000263	 \\
747	& 	2456460.804500	 & 	0.000243	 \\
747	& 	2456460.805467	 & 	0.000643	 \\
782	& 	2456510.868182	 & 	0.000602	 \\
782	& 	2456510.866993	 & 	0.000150	 \\
786	& 	2456516.586674	 & 	0.001191	 \\
789	& 	2456520.880123	 & 	0.000637	 \\
798	& 	2456533.752605	 & 	0.000707	 \\
798	& 	2456533.754796	 & 	0.000152	 \\
828	& 	2456576.662886	 & 	0.001087	 \\
837	& 	2456589.541970	 & 	0.000899	 \\
851	& 	2456609.566526	 & 	0.000426	 \\
1012	& 	2456839.854400	 & 	0.001226	 \\
1028	& 	2456862.740854	 & 	0.000482	 \\
1042	& 	2456882.765657	 & 	0.000726	 \\
1044	& 	2456885.624290	 & 	0.000529	 \\
1049$^{*}$	& 	2456892.778598	 & 	0.000957	 \\
1065	& 	2456915.660396	 & 	0.001232	 \\
1084	& 	2456942.838802	 & 	0.000784	 \\
1088	& 	2456948.563838	 & 	0.000740	 \\
1316	& 	2457274.684578	 & 	0.001838	 \\
1330	& 	2457294.708862	 & 	0.001400	 \\
1509	& 	2457550.747972	 & 	0.000306	 \\
1539	& 	2457593.656923	 &	0.000242	 \\
1544	& 	2457600.809853	 & 	0.000393	 \\
1551	& 	2457610.822857	 & 	0.000203	 \\
\hline                                   
\end{tabular}

Note: Asterisks indicate the transits excluded from the calculation of ephemeris.
\end{table}

\begin{figure}
   \centering
   \includegraphics[width=.4\textwidth]{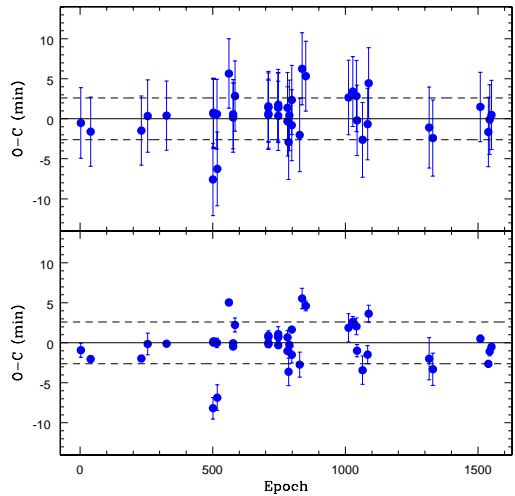}
      \caption{O-C data-points versus Epoch considering the 45 light curves without anomalies in their transits. The O-C values are the observed mid-transit times minus the mid-transit times predicted using a specific ephemeris equation, while Epoch represents the number of transits since the minimum reference time shown in the same equation. Bottom panel: O-C values obtained from the ephemeris given by Eq. (3). Upper panel: O-C values obtained from the ephemeris given by Eq. (5). Here, the error bars include the extra variance component of 4.32 minutes, added in quadrature to the mid-transit times errors, which was computed from a maximum-likelihood approach. In both panels, dashed lines indicate $\pm \sigma$ ($\sigma = 2.61$ minutes) which represents the standard deviation of the data.}
         \label{FigVibStab}
   \end{figure}

\begin{figure}
   \centering
   \includegraphics[width=.4\textwidth]{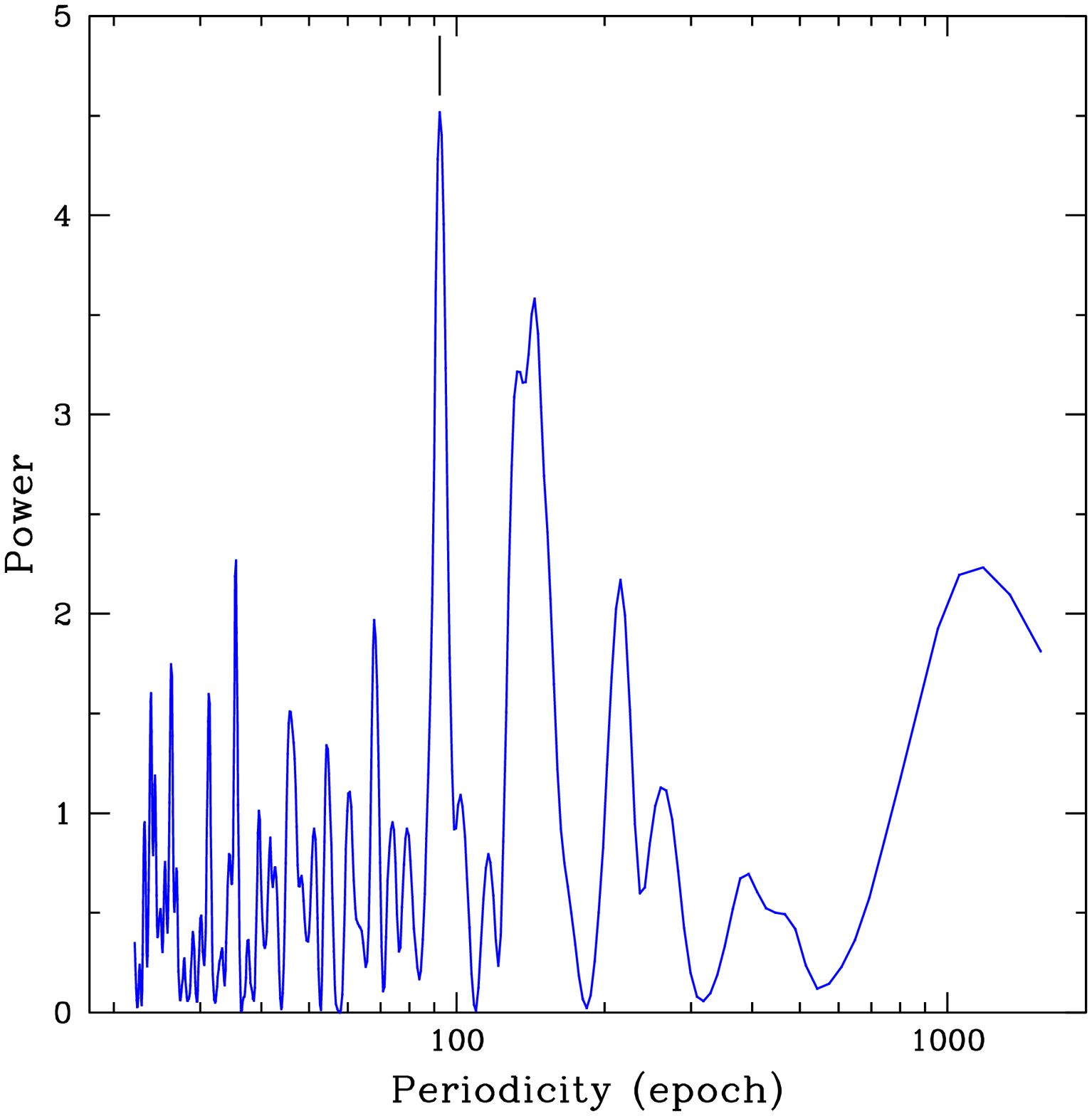}
      \caption{Lomb-Scargle periodogram of the O-C values corresponding to the 45 analysed light curves (continuous blue line). The most significant peak (P = 92.4 epochs) with FAP = 42$\%$ is marked. Here, ``epoch'' represents the number of transits since the minimum reference time shown in Eq. (5), while FAP (False Alarm Probability) is the probability that random noise produces a peak with power similar or higher than the one of the most significant peak in a certain period range. In this case, FAP was estimated through 10000 Monte Carlo simulations.}
         \label{FigVibStab}
   \end{figure}

We have excluded 4 light curves from the ephemeris computation. The transit observed during the night of 2014 August 22 was not considered because it is partial, while those transits acquired the nights of 2011 August 31, 2012 July 26, and the one obtained in the Sloan $z'$ filter the night of 2012 July 2 were not included because they show some visible anomalies during the transit. We suspect that these asymmetries could be caused by stellar activity or simply by bad weather conditions (thin clouds, fog, etc.). We computed new ephemeris fitting a linear model to the remaining 45 light curves through least-squares weighted by the mid-transit times uncertainties. Considering these data we obtained,

\begin{equation}
T_{\mathrm{0}}(E)=2455392.31738(36) BJD_\mathrm{TDB} +  E \times 1.43037126(50)
\end{equation}

\noindent where $E$ represents the epoch, i.e., the number of transits since the minimum reference time. The errors for the orbital period and minimum reference time correspond to the last digits and were computed from the covariance matrix of the fit. We obtained a $\chi^{2}_{\mathrm{r}} = 13.17$ which implies that a linear ephemeris does not properly represent the mid-transit times behaviour. In the bottom panel of Fig. 4 we plot O-C versus Epoch, where the O-C values shown in the Y-axis are the observed mid-transit times $T_\mathrm{0}$ minus the ones predicted using the ephemeris given by Eq. (3). Here, dashed lines represent $\pm \sigma$, i.e., the standard deviation of the sample ($\sigma = 2.61$ minutes). In this case, it can be seen that the O-C data-points show some substantial scatter, probably due to the magnetic activity of the host-star. To fully account for the stellar activity influence or any other remaining uncorrelated noise, we introduced an additional variance component ($\sigma_{s}$) in the ephemeris calculation. We estimated the value of this extra contribution using the same approach employed by \citet[see their Section 4.1]{haywood}, who followed a procedure similar to the one described in \citet{collier}. This procedure basically consists in determining, through an iterative process, the value of $\sigma_{s}$ that maximises the likelihood ($\mathcal{L}$),

\begin{equation}
\mathrm{ln}(\mathcal{L})=-\frac{n}{2} \mathrm{ln}(2\pi)-\frac{1}{2}\chi^{2} -\frac{1}{2}\sum_{i=1}^{n}\mathrm{ln}(e^{2}_{\mathrm{T_{\mathrm{0}}}}+\sigma^{2}_{s}),
\end{equation}

\noindent where $\chi^{2}$ is the chi-squared value of the $n =$ 45 data-points with uncertainties $e_{\mathrm{T_{\mathrm{0}}}}$. Through this approach, we estimated an extra variance component of 4.32 minutes, which was added in quadrature to the mid-transit times errors, and the following linear ephemeris,

\begin{equation}
T_{\mathrm{0}}(E)=2455392.3170(4) BJD_\mathrm{TDB} +  E \times 1.43037148(53).
\end{equation}

\noindent In this case, we obtained $\chi^{2}_{\mathrm{r}} = 0.32$. The O-C values computed from Eq. (5) are shown in the upper panel of Fig. 4. Here, it is possible to observe that considering an additional variance contribution of 4.32 minutes, the O-C data-points are within the level of the error bars. To search for a periodicity in these data we ran two different tasks to the O-C values: a Lomb-Scargle (LS) periodogram \citep{horne} and a Phase Dispersion Minimization (PDM) algorithm \citep{pdm} provided by IRAF. Both routines find very similar peaks at P = 92.4 epochs with a FAP\footnote{FAP, for False Alarm Probability, is the probability that random noise produces a peak with power similar or higher than the one of the most significant peak in a certain period range. In this case, FAP was estimated through 10000 Monte Carlo simulations.} of 42$\%$ for the LS periodogram (Fig. 5), and P = 91.9 epochs with $\Theta =$ 0.835 for the PDM algorithm\footnote{$\Theta$ is a statistic that indicates how significant is the value found for a certain period. It is defined as $\Theta = s^2/\sigma^2$ where $\sigma^2$ is the variance of the analysed data series and $s^2$ is computed from the variances of data subsets obtained by splitting the original data series into several sub samples. If the found period is not true then $\Theta \sim$ 1, while $\Theta \sim$ 0 when the found period is correct. Note that $\Theta \sim$ 1 and $\Theta \sim$ 0 are equivalent to large and small values of FAP, respectively.}. These high values for the false alarm probabilities and the absence of a clear periodic behaviour in the O-C data-points seem to indicate that the period found around 92 epochs is not real. However, in the upper panel of Fig. 4 it is clear that some points strongly deviate from the predicted mid-transit times. We explored the causes of these departures looking for possible correlations between the O-C values and several indicators of the light curve quality.

\begin{figure*}
  \centering
 \includegraphics[width=.8\textwidth]{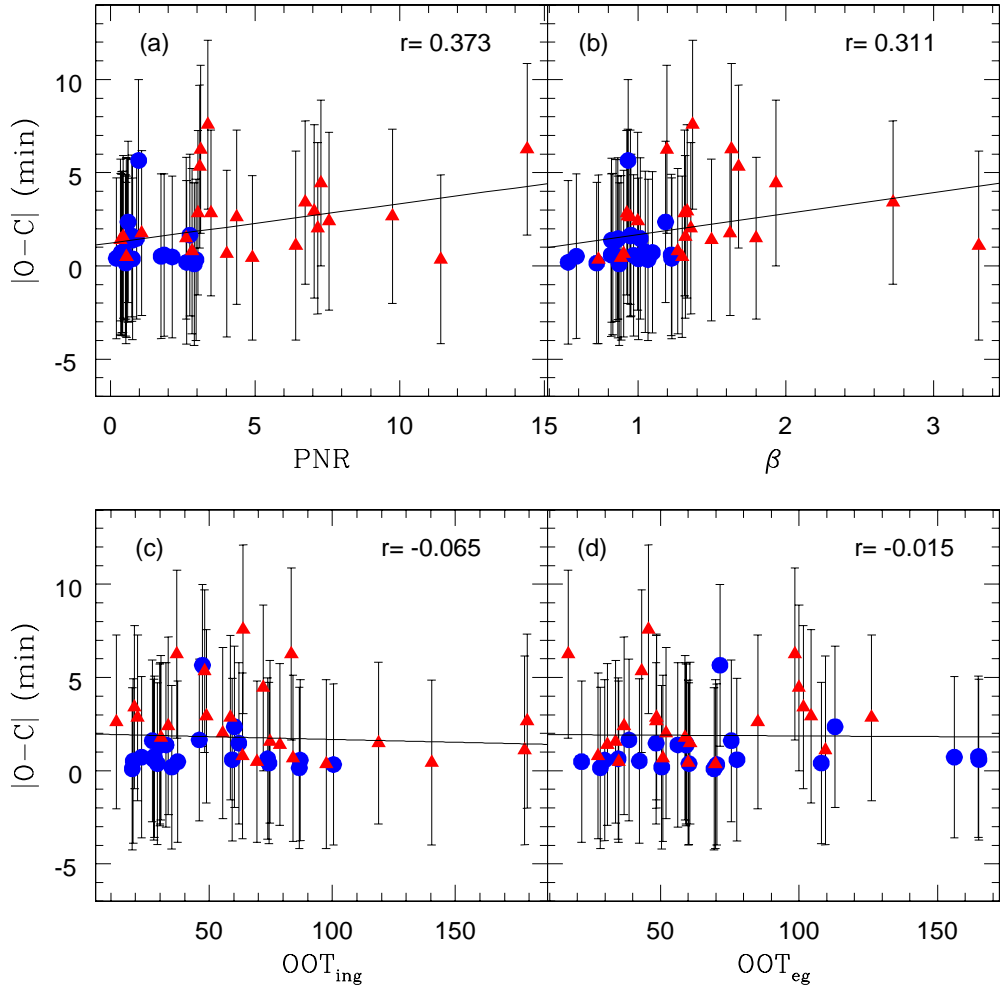}
   \caption{|O-C| data-points versus photometric noise rate (PNR), red noise ($\beta$), time before ingress (OOT$_{ing}$), and time after egress (OOT$_{eg}$). Here PNR and $\beta$ denote photometric accuracy of the light curves.  Blue circles indicate high quality transits with PNR $\le$ 3 and $\beta$ $\le$ 1.25, while red triangles correspond to those poor quality light curves with PNR > 3 and/or $\beta$ > 1.25. Solid lines represent the best-fitting to the data and the $r$ parameters are the correlation coefficients. We regarded -1 < r $\le$ -0.8 or 0.8 $\le$ r < 1 as strongly, -0.8 < r < -0.5 or 0.5 < r < 0.8 moderatly, and -0.5 $\le$ r < 0 or 0 < r $\le$ 0.5 weakly (positive or negative) correlated parameters. Taking this classification into account, no strong correlations between the |O-C| values and PNR, $\beta$, OOT$_{ing}$, and OOT$_{eg}$ are observed.}
     \label{FigVibStab}
\end{figure*}

In Fig. 6, we plot the absolute values of O-C versus PNR (Fig. 6a) and $\beta$ (Fig. 6b). \citet{gibson} have demonstrated that the duration of the observations before/after the transit ingress/egress might play an important role in the normalization of the light curve and therefore in the determination of $T_\mathrm{0}$. Bearing this in mind, we also plot in Figs. 6c and 6d, the |O-C| values versus
the duration of the out-of-transit observations before ingress (OOT$_{ing}$)  and the out-of-transit observations after egress (OOT$_{eg}$), respectively. In all the plots we distinguish between transits with PNR $\le$ 3 and $\beta$ $\le$ 1.25 (blue circles) and those with PNR > 3 and/or $\beta$ > 1.25 (red triangles). The 4 light curves excluded from the ephemeris computation were also not considered in this analysis. For each figure, we performed a linear fit (solid line) to the data and computed the correlation coefficient $r$. We regarded -1 < r $\le$ -0.8 or 0.8 $\le$ r < 1 as strongly, -0.8 < r < -0.5 or 0.5 < r < 0.8 moderatly, and -0.5 $\le$ r < 0 or 0 < r $\le$ 0.5 weakly (positive or negative) correlated parameters. Taking this classification into account, no strong correlations between the absolute values of O-C and PNR, $\beta$, OOT$_{ing}$, and OOT$_{eg}$ are observed. 

However, it is interesting to note that, with the exception of only one point, which corresponds to the epoch 561, the |O-C| data-points from the best quality transits (blue circles) do not exceed 2.4 minutes, whereas those derived from light curves of less quality (red triangles) show deviations up to 7.6 minutes. Moreover, mid-transit times errors are, on average, 4.34 and 4.51 minutes for the blue circles and the red triangles, respectively. These results would imply a possible connection between the transit quality and the calculated value of $T_\mathrm{0}$ and its uncertainty. We investigated this possibility by plotting in Figs. 7a and 7b the errors in $T_\mathrm{0}$ ($e_\mathrm{T_{0}}$) versus PNR and $\beta$, respectively. In Fig. 7b we obtained a coefficient $r$ of 0.447, which suggests a weak correlation between $e_\mathrm{T_{0}}$ and red noise. However, this correlation is fully dependent on the data-point with a $\beta$ factor of 3.3, which corresponds to the epoch 1316. If this transit is removed, $r$ decreases to 0.062, showing almost no correlation between both quantities. On the other hand, in Fig. 7a we show with a solid line the best linear fit to the data-points. In this case, we found a moderate correlation ($r = 0.672$) between the errors in the determination of the mid-transit times and PNR, suggesting that less quality light curves provide less accurate values of $T_\mathrm{0}$.

\begin{figure*}
   \centering
   \includegraphics[width=.8\textwidth]{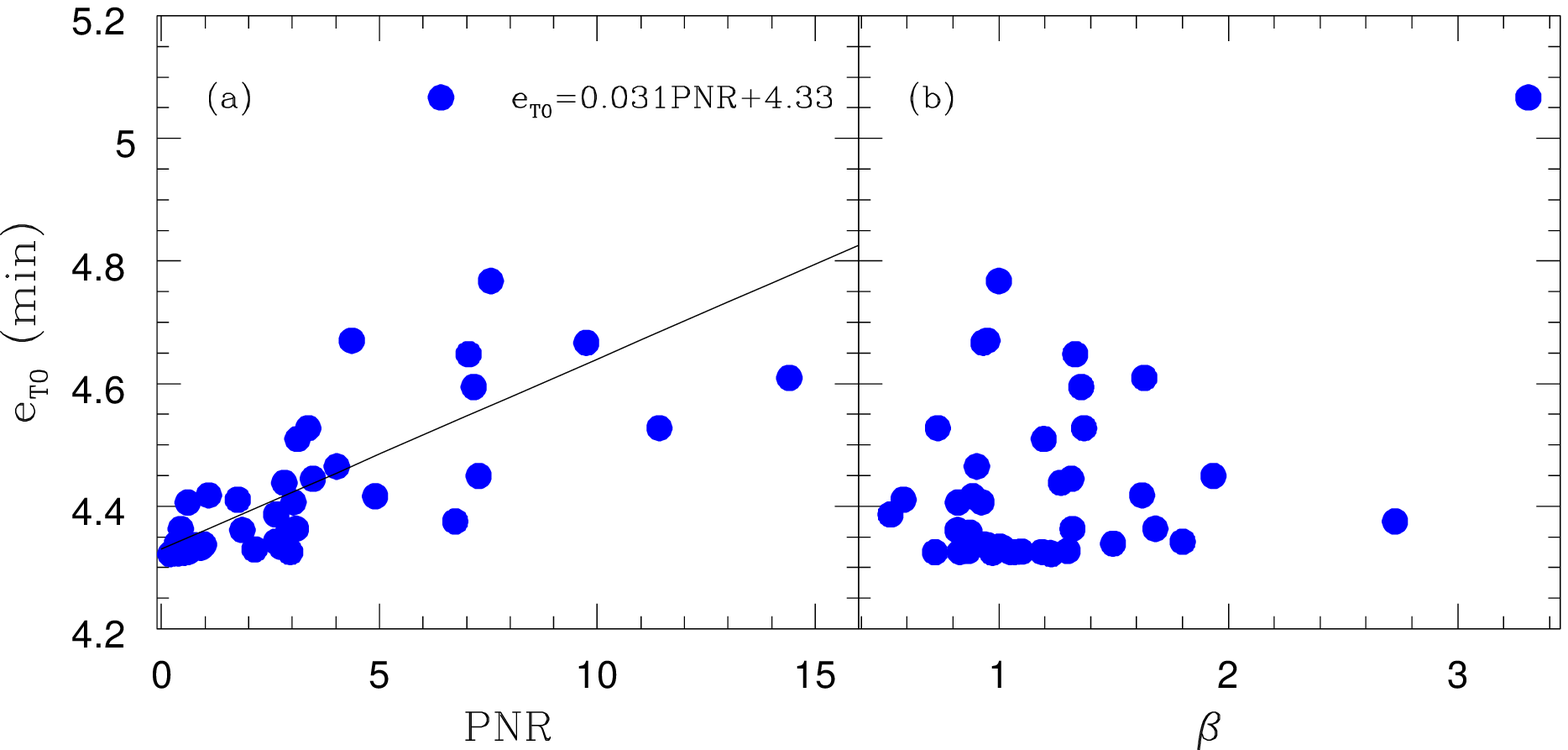}
      \caption{Uncertainties in mid-transit times ($e_\mathrm{T_{0}}$) versus PNR (panel a) and $\beta$ (panel b) factors. The black solid line represents the best-fitting to the data-points whose equation is indicated in the figure, and the $r$ parameter is the correlation coefficient. For the data in panel b, we obtain $r = 0.447$ which suggests a weak correlation between $e_\mathrm{T_{0}}$ and red noise. However, this correlation is fully dependent on the data-point with $\beta = 3.3$, which corresponds to the epoch 1316. If this transit is removed, $r$ decreases to 0.062 showing almost no correlation between both quantities. On the other hand, for the data in panel a, we find a positive moderate correlation ($r = 0.672$) between the errors in the determination of the mid-transit times and PNR, suggesting that less quality light curves provide less accurate values of $T_{0}$.}
         \label{FigVibStab}
   \end{figure*}

Considering these results, we decided to re-calculate the ephemeris including only those light curves with mid-transit times uncertainties smaller than 1 minute according to the values presented in Table 7 and no visible anomalies (Fig. 8). 
This reduced our sample to 31 transits spanning 6 complete years. By performing a maximum-likelihood analysis similar to that applied for determining Eq. (5), we obtained

\begin{equation}
T_{\mathrm{0}}(E)=2455392.3176(2) BJD_\mathrm{TDB} +  E \times 1.43037123(26),
\end{equation}

\noindent with $\chi^{2}_{\mathrm{r}}=0.83$. In this case, the estimated extra variance component added in quadrature to the errors in $T_{\mathrm{0}}$, was 1.7 minutes.
Contrary to the results obtained taking the 45 transits into account, an LS periodogram of the 31 data-points does not show any significant peak, which confirms that the period of 92 epochs determined before is not   
real. This is in agreement with the finding by \citet{ciceri} who did not find any periodic signal in the data. We also investigated if the use of different filters might affect the measurement of ephemerides. For the light curves corresponding to the nights of 2012 October 16 and 2013 April 24 observed with the GROND $z'$ and $g'$ filters, respectively, we repeated the fitting procedures explained in Section 3.2 and at the beginning of this section. However, this time, we took as initial values for the limb-darkening coefficients those corresponding to a different filter (the Johnson V filter). Then, we used the values and errors of the mid-transit times obtained from these fits and re-computed ephemeris, which was compared with that presented in Eq. (3). Given that the result of this comparison shows that the change in the ephemeris is within its 1$\sigma$ error bars, we conclude that the ephemeris, and hence the measured mid-transit times, are not affected by the filters used to carry out the observations.


On the other hand, it is known that WASP-46 is an active star with a rotation period of 16 days, determined from the photometric variations produced by spots of magnetic origin \citep{anderson}. 
Several works \citep{oshagh, ioannidis} caution that asymmetries in the light curves due to the passage of the planet in front of one or several spots during transit may lead to measure mid-transit times that strongly deviate from the predicted ones. Therefore, some of the outliers observed in Fig. 8, which deviate more than $\sigma = 1.66$ minutes from the predictions, can be due to anomalies during the transit produced by the presence of unseen stellar spots. 

\begin{figure}
   \centering
   \includegraphics[width=.4\textwidth]{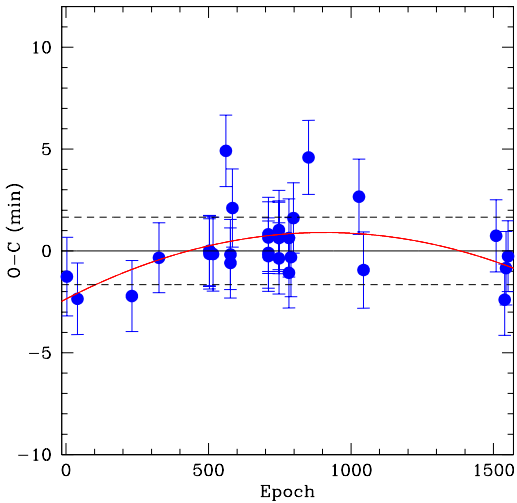}
      \caption{O-C data-points versus Epoch. Here, O-C values are the observed mid-transit times minus the mid-transit times predicted using the ephemeris given by Eq. (6), while Epoch represents the number of transits since the minimum reference time shown in the same equation. We only plot the 31 light curves with uncertainties in their mid-transit times smaller than 1 minute according to the values presented in Table 7. Dashed lines indicate $\pm \sigma$ ($\sigma = 1.66$ minutes) which represents the standard deviation of the data, while the red continuous line is the quadratic fit obtained from Eq. (11). Error bars include an additional variance component of 1.7 minutes, added in quadrature to the mid-transit times errors, which was computed from a maximum-likelihood approach.}
         \label{FigVibStab}
   \end{figure}


For completeness, we assessed if our standard deviation in the O-C data-points might be compatible with the amplitude of the TTV expected by the Applegate effect\footnote{The Applegate effect is a mechanism that produces changes in the orbital period of the components of a binary system, due to quasi-periodic variations in the stellar quadrupole moment caused by magnetic activity cycles. This same effect is also expected to occur in planetary systems when the host-star is magnetically active \citep{watson}.} \citep{applegate}. According to Eq. (13) of \citet{watson}, this amplitude would be of less than 1 second in 6 years, which is far below the $\sigma \sim$ 1.6 minutes found in our data. Hence, we can discard the  variations in the quadrupole moment of WASP-46  as the cause of the observed dispersion in our mid-transit times. 

Finally, using Eq. (33) of \citet{agol} with the TTV dispersion determined for the 31 transits with
$e_\mathrm{T_{0}}$ less than 1 minute, we found that it is possible to exclude perturbers with masses larger than 2.3, 4.6, 7, and 9.3 $M_{\mathrm{\earth}}$ located in the positions of the first-order mean-motion resonances 2:1, 3:2, 4:3, and 5:4 with WASP-46b, respectively.

\subsection{Analysis of long-term variations in transit depth and orbital inclination}

The presence of a third body (exomoon, ring, or another planet) in the system can also cause periodic variations in depth and/or
orbital inclination. To study this possibility we analysed the long-term behaviour of  
the photometric parameters $k$ and $i$ as a function of time for the 45 light curves without visible anomalies in their transits. To pursue this aim, we ran the JKTEBOP code individually on each light curve considering the depth, the scale factor, and the coefficients of the polynomial to fit the out-of-transit data-points as free parameters, while the remaining ones were fixed to the values obtained in Section 3.2. We repeated the same procedure for the orbital inclination but, in this case, the sum of the fractional radii was also allowed to vary because $i$ and $\Sigma$ are correlated parameters. As before, we excluded from this study the incomplete light curve and the 3 light curves showing visible anomalies during the transit. In Fig. 9 we present our results. In both cases, we ran an LS periodogram and no significant peak was found in the data. However, for the $k$ parameter some transits present depth values that depart from the standard deviation of the sample (here $\sigma = 0.0104$), similar to what we obtained for the mid-transit times. We suspect that these departures are consequence of the stellar activity present in WASP-46 (see e.g. \citealt{czesla, croll}).
Since our sample consists of light curves observed in 11 different filters, including wavelengths from 477 to around 914 nm, we extended the analysis performed by \citet{ciceri} and explored the behaviour of the planetary radius as a function of wavelength. With this purpose, the 45 measurements of $k$ were grouped together according to the filter in which they were observed. Then, for each dataset we computed a weighted average of $k$ and adopted as error its standard deviation. These values are shown in Table 8. By fitting a linear model through weighted least-squares to this data, we obtained a slope $m = -0.98 \times 10^{-5}$ which is fully consistent with the value found by \citet{ciceri}. Assuming a scaling law for the cross section of the dominant species given by $\sigma=\sigma_{0}(\lambda/\lambda_{0})^{\gamma}$ \citep{lecavelier}, the slope \textit{m} will depend on the planet's atmospheric properties as follows:

\begin{equation}
m=\frac{\gamma T'_{eq} k_{B}}{\mu g_{P}},
\end{equation}

\noindent where $k_{B}$ is the Boltzmann constant, $\mu$ the mean molecular weight, and $T'_{eq}$ and $g_{P}$ are the previously defined planet's modified equilibrium temperature and planetary surface gravity, respectively. Through the measured value of \textit{m}, it is possible to determine if a scattering process is taking place in the planet's atmosphere and to infer which chemical component is producing it. However, we will not perform this kind of analysis because it is beyond the scope of this paper. In Fig. 10 we show the measured values of \textit{k} and their errors as a function of wavelength. The linear model that best fits the data is indicated by a continuous black line. Furthermore, the upper panel of Fig. 9 shows that all the values of $i$ agree, considering the errors, with the mean value measured for the orbital inclination.

\begin{table}
\caption{Average $k$ values obtained for each filter}             
\label{table:8}      
\centering          
\begin{tabular}{lccc}     
\hline\hline       
Filter & $\lambda_{C}$ (nm)	&	$k$ &	 $\sigma_{k}$\\
\hline                    
GROND $g'$	& 477.0 &	0.1419 & 	0.0026\\
Gunn $g'$ & 	516.9 & 0.1452 & 0.0040\\
Jhonson $V$ & 551.0 & 0.1287 & 0.0125\\
clear & 592.7 & 0.1361 & 0.0034\\
GROND $r'$ &	623.1 & 0.1442 & 0.0023\\
Jhonson $R$ &	640.7 & 0.1369 & 0.0039\\
Bessell $R$	 & 648.9 & 0.1369 & 0.0043\\
Gunn $r'$	& 664.1 & 0.1449 & 0.0048\\
GROND $i'$ & 762.5 & 0.1317 & 0.0040\\
Cousins I$+$Sloan $z'$ & 849.9 & 	0.1306 & 0.0086\\
GROND $z'$	& 913.4 & 0.1408 & 0.0031\\
\hline                  
\end{tabular}

Note: The $\lambda_{C}$ values for the passbands ``clear" and ``Cousins I$+$Sloan $z'$" are the average $\lambda_{C}$ of the Johnson $V$ and Cousins $R$ filters, and the Cousins $I$ and the Sloan $z'$ filters, respectively. 
\end{table}

\begin{figure*}
   \centering
   \includegraphics[width=.6\textwidth]{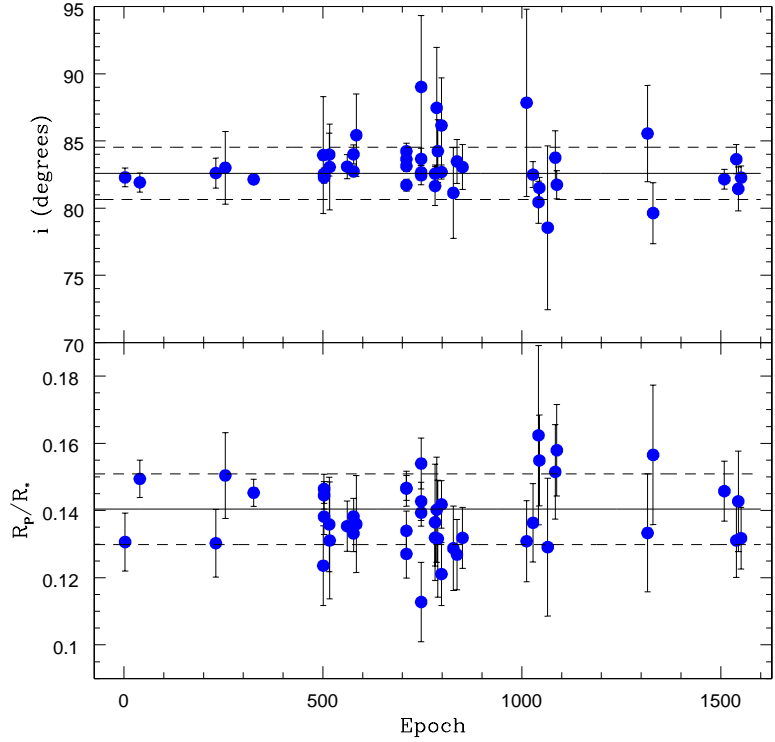}
      \caption{Long-term variations of $i$ (\textit{upper panel}) and $k$ (\textit{lower panel}). Here, Epoch represents the number of transits since the minimum reference time given by Eq. (4). Black solid lines indicate the weighted averages of the sample and dashed lines indicate $\pm \sigma$. Error bars are also shown.}
         \label{FigVibStab}
   \end{figure*}

\begin{figure}
   \centering
   \includegraphics[width=.45\textwidth]{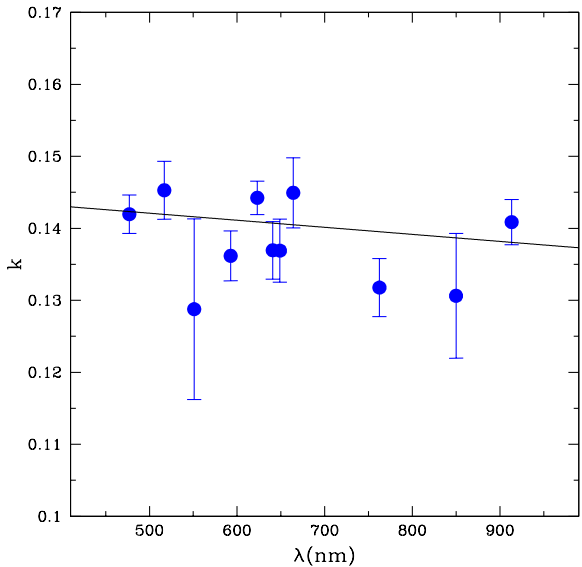}
      \caption{Variation of the transit depth of WASP-46b as a function of wavelength. Blue points are the average \textit{k} values shown in Table 8 with their respective error bars. The black continuous line represents the linear model that best fits the data with a slope $m = -0.98 \times 10^{-5}$.}
         \label{FigVibStab}
   \end{figure}

\subsection{Searching for a possible orbital decay}

Considering the close proximity of WASP-46b to its host-star (a $\sim$ 0.024 AU), it is interesting to study the possibility of orbital decay. According to \citet{matsumura} planetary systems are ``Darwin unstable'' (i.e., they have no tidal equilibrium states) when the ratio of the total angular momentum of the system (L$_{tot}$) to some critical value (L$_{crit}$) is lower than  1.  Here,

\begin{equation}
L_{tot}=L_{orb}+(C_\mathrm{\star}+C_\mathrm{P})n=\frac{M_\mathrm{\star} M_\mathrm{P}}{\sqrt{M_\mathrm{\star} + M_\mathrm{P}}}\sqrt{Ga(1-e^2)} +(C_\mathrm{\star}+C_\mathrm{P})n
\end{equation}

\noindent and,

\begin{equation}
L_{crit}=4 \left( \frac{G^2}{27}\frac{M^{3}_\mathrm{\star} M^{3}_\mathrm{P}}{M_\mathrm{\star} + M_\mathrm{P}}(C_\mathrm{\star}+C_\mathrm{P}) \right) ^{\frac{1}{4}},
\end{equation}

\noindent where $L_{orb}$ is the orbital angular momentum, $G$ the gravitational constant, $n = 2\pi/P$ the mean motion, and $C = \alpha M R^2$ the moment of inertia.
Using Eqs. (8) and (9) for the system under study, we calculated L$_{tot}$/L$_{crit}$ $\sim$ 0.11\footnote{This value was computed considering the planet and the star as point masses ($\alpha = 1$ in both cases).} which means that the final fate of WASP-46b is to eventually fall onto the stellar surface. In this case, since $e$ is nearly zero and the orbit is supposed to be synchronised, the tidal forces acting on the exoplanet can be considered negligible. However, since the stellar rotation period is larger than the orbital period, it is expected that tides continue to act on the host-star, decreasing the semimajor axis until the planet reaches its Roche limit ($a_\mathrm{R}$) and is tidally disrupted \citep{penev}. 
According to \citet{faber}, the critical separation from which the planet starts to lose mass via its Roche lobe, $a_\mathrm{R}$, is given by, 

\begin{equation}
R_{\mathrm{P}}=0.462q^{1/3}a_{\mathrm{R}}
\end{equation}

\noindent where $q = M_{P}/M_{\star}$ is supposed $\ll$ 1. This relation between $R_\mathrm{P}$ and $a_\mathrm{R}$ is based on the Roche lobe radius determined by \citet{pac}, who considered the classical stellar two-body problem supposing both stars as point masses in a circular orbit. In our case, adopting the stellar and planetary masses and the planet radius computed in Section 3.3, we found $a_{R} = 0.00972$ AU.  Assuming the current value of $a = 0.02407$ AU determined in this work, this result implies that WASP-46b has not crossed its Roche limit yet. However, we cannot rule out the possibility that the planet is losing mass through evaporation due to stellar radiation, given that this mechanism is not contemplated in the Roche limit calculation.

Orbital decay manifests as a systematic decrease of the orbital period. This implies that successive transits begin at times earlier than predicted ones, and therefore O-C values become systematically negative. The usual method to search for this shortening in the orbital period is to fit the mid-transit times with a quadratic and a linear ephemeris and compare which of both models better represents the data. 

In this particular case, we performed a maximum-likelihood analysis to fit the mid-transit times corresponding to the 31 light curves for which $e_{\mathrm{T_{\mathrm{0}}}} <$ 1 minute with the quadratic ephemeris equation of \citet{adams},  

\begin{equation}
T_{\mathrm{0}}(E)= T_{\mathrm{minref}} +  E \times P + \delta P \times \frac{E(E-1)}{2}
\end{equation}

\noindent where $T_{\mathrm{minref}}$ is the reference minimum time and $\delta P=P \times \dot{P}$. In Fig. 8, the red continuous line represents the quadratic fit to the data. We obtained a variation of the orbital period per epoch ($\delta P$) of 
$(-5.41 \pm 2.25) \times 10^{-9}$ days and  consequently a variation of the orbital period per year ($\dot{P}$) of
$-0.119 \pm 0.049$ s yr$^{-1}$. These small values for $\delta P$ and $\dot{P}$ would be  consistent with a constant orbital period. Following the methodology used in previous studies (see e.g. \citealt{chen14, hoyer113}), we applied to both the linear and the quadratic models the Bayesian Information Criterion (\textit{BIC}) defined as,

\begin{equation}
BIC= \chi^{2} + k_\mathrm{FP} \ln N,
\end{equation}

\noindent where $k_\mathrm{FP}$ is the number of free parameters for the adjustment and $N$ is the number of data-points. \textit{BIC} is a useful tool to evaluate which of the fits better represents the data. In this sense, the best-fitting will be the one with the lowest value of \textit{BIC}. 
For WASP-46b we obtained $BIC = 33$ for the linear model and $BIC = 28$ with a $\chi^{2}_{\mathrm{r}} = 0.56$ for the quadratic fit. These results indicate that a quadratic ephemeris is a better representation of the mid-transit times as a function of epoch. However, although we cannot rule out a possible slow decreasing rate of the period, we caution the reader that this evidence is still not significant and more observations are required before we can make conclusive statements on the orbital decay of the system.\\

Although the analysis presented above did not reveal any periodic variation in the mid-transit times or a significant shortening in the orbital period, the information obtained is useful to determine a lower limit on the tidal quality factor Q$_{\star}$. This parameter is related to the rate of tidal dissipation within the host-star and hence has an influence on the rate of orbital decay \citep{penev}. Larger values of Q$_{\star}$ imply slower orbital evolutions. This points out the importance of this parameter in the theories looking to explain the formation of close in planets and their subsequent dynamic evolution. However, in spite of the theoretical effort put on the determination of Q$_{\star}$ (see e.g. \citealt{matsumura, penev}), the mechanisms of tidal dissipation in planets and stars are not well understood yet. Recently, several groups \citep{croll, hoyer43, hoyer113, macie} have started to estimate the values of the tidal quality factor through observations of individual systems. We followed the procedure applied in these previous works and using a 1$\sigma$ value based on our measured orbital decay, we adopted -0.07 s yr$^{-1}$ as the upper limit of the estimated value for $\dot{P}$. Based on this number,  we calculated Q$_{\star} > 7 \times 10^{3}$ for WASP-46 from Eq. (5) of \citet{birkby}. Although this value for the tidal quality factor is much lower than those normally assumed (between $10^{5}$ - $10^{10}$), it is physically plausible given that small values of Q$_{\star}$ have been also obtained in previous studies (see e.g. \citealt{adams, blecic}).
Finally, we used the value of the lower limit determined for Q$_{\star}$ to compute a lower limit for the remaining lifetime of $>$ 0.34 Myr before WASP-46b falls onto the star.

\section{Summary and conclusions}

In this work we present 12 new transits of WASP-46b observed between 2012 and 2016. We use these observations and another 37 light curves collected from previous works and the Exoplanet Transit Database to re-determine the system's parameters and to compute new ephemeris. 
From the complete (full transit coverage) and higher quality light curves we estimate photometric parameters which are in full agreement with the values published by \citet{anderson}, but are slightly different from those
calculated by \citet{kjurk} and \citet{ciceri}, with the exception of $k$ which is coincident in this last case. For the physical parameters, we find a value for the planetary radius consistent with the one computed in the discovery paper and somewhat larger than the one measured by \citet{ciceri}. Then, the main result of the first part of this work is that our estimations suggest a value for the planetary density between previous determinations by \citet{anderson} and \citet{ciceri}.

We also perform the first homogeneous TTV study for this system over 6 years of observations. For the 45 O-C data-points corresponding to those light curves without visible anomalies during the transit, we get a dispersion of 2.61 minutes although we do not find any periodicity. To explain this high value of $\sigma$, we search for possible correlations between |O-C| and PNR, $\beta$, and the duration of the out-of-transit observations before ingress and after egress, but no significant correlation is detected. However, two interesting results arise from this analysis. First, best quality light curves (complete transits with PNR $\le$ 3 and $\beta$ $\le$ 1.25), with the exception of epoch 561, show |O-C| values up to 2.4 minutes with an average error of 4.34 minutes, while transits of poor quality present |O-C| values as much as 7.6 minutes with a mean error of 4.51 minutes. Second, we find a moderate correlation ($r = 0.672$) between the errors in the mid-transit times and PNR. This finding agrees with the results obtained in previous works and shows that poor quality transits imply not only less accurate values of $T_\mathrm{0}$, but also larger errors. Since ephemerides are generally computed through weighted least-squares, the values of $P$ and reference minimum time are only slightly affected by poor quality light curves. However, these poor quality transits (often used in TTV's studies based on observations acquired with ground-based facilities) could mimic the variations in the O-C data-points produced by another body. Therefore, caution must be taken when these low quality observations are included in this kind of analysis. 

Given that we showed that low-quality data usually provide less accurate values of mid-transit times with errors often underestimated, to calculate the ephemeris we only consider the 31 complete light curves with errors in $T_\mathrm{0}$ smaller than 1  minute.
In this case, we find a standard deviation of 1.66 minutes for the O-C data-points, which is significantly smaller than the one obtained considering all the 45 data-points. 
Since no periodic variation is detected in the data, we exclude the possibility that a second body gravitationally bound to the system can explain this result. We also discard the Applegate effect as a possible cause, since the amplitude of the variations produced by this phenomenon (less than 1 second) is significantly lower than the dispersion found ($\sim$ 1.6 minute). Alternatively, the high standard deviation we obtain can be due to stellar activity.
In addition, our TTV dispersion allows us to exclude bodies with masses larger than 2.3, 4.6, 7, and 9.3 $M_{\mathrm{\earth}}$ at the first-order mean-motion resonances 2:1, 3:2, 4:3, and 5:4 with WASP-46b, respectively. Moreover, we do not detect any periodic behaviour in depth and orbital inclination for the 45 light curves. Several values of $k$ differ more than $\pm \sigma$ from the mean value, probably due to the effect of unseen stellar spots in the light curves. 

Given the short distance between WASP-46b and the star, we also search for a possible orbital decay. Through the computation of the total angular momentum of the system, we conclude that the planetary orbit is unstable and WASP-46b will eventually spiral in toward its host-star. We also estimate that the planet has not crossed its Roche limit yet. 
Furthermore, we find that a quadratic fit to the 31 best mid-transit times is a better (BIC = 28) representation of the data than the linear model (BIC = 33), which prevents us from ruling out the possibility that the orbital period of WASP-46b might be decreasing. However, from the quadratic model we estimated small values for $\delta P = (-5.41 \pm 2.25) \times 10^{-9}$ days and $-0.119 \pm 0.049$ s yr$^{-1}$, suggesting that if a decay in the planetary orbit is actually taking place, the variation rate of the period is very low. Moreover, it is important to mention that even considering a typical light curve precision of $2 \times 10^{-3}$ and 6 years of observations, our results cannot significantly demonstrate a slow decrease of the orbital period of WASP-46b. Hence, this trend is not conclusive and needs to be confirmed by extending the baseline of transit observations.
This value of $\dot{P}$ allows us to compute a lower limit on the tidal dissipation coefficient of Q$_{\star} > 7 \times 10^{3}$. On the other hand, according to Eq. (7) of \citet{birkby} and assuming our current light curve precision and O-C dispersion of 1.6 minutes as the expected transit time shift, we would be able to rule out values of Q$_{\star} < 10^{5}$ after 6 additional years of transit observations.
Given that there is still no clear evidence to decide if the orbit of the planet is decaying or not, the transfer of angular momentum from the planetary orbit to the stellar surface proposed by Maxted et al. (2015) remains a possibility that might explain the discrepancy between the gyrochronological and isochronal ages.

 \section*{Acknowledgements}
\addcontentsline{toc}{section}{Acknowledgements}

R. P. and E. J. acknowledge the financial support from CONICET in the form of postdoctoral fellowships. R. P. also thanks to Dr. Simona Ciceri for kindly providing the light curves of her work. R. P. and E. J. are also grateful to the operators of the 1.54-m telescope at EABA, Cecilia Qui\~nones and Luis Tapia, for their support during the observing runs, and to Estefan\'ia Vendemmia for observing the transit of 2014, August 22. R. P. thanks Martin Ma\v{s}ek for nicely providing information about the observations of WASP-46b published in the ETD. The authors acknowledge support from the PIP 2013-2015 GI 11220120100497 of CONICET (Argentina). This research has made use of the SIMBAD database, operated at CDS, Strasbourg, France and NASA's Astrophysics Data System. We also thank the referee for a thorough review of the manuscript and constructive comments and suggestions, which improved the content and quality of the paper.




\bibliographystyle{mnras}
\bibliography{wasp46}



\bsp	
\label{lastpage}
\end{document}